\documentclass[preprint,12pt]{aastex62}

\received{\today}
\revised{\today}
\accepted{\today}

\shorttitle{S-Type Planet Stability}
\shortauthors{Quarles, Li, Kostov, \& Haghighipour}

\begin{document}

\title{Orbital Stability of Circumstellar Planets in Binary Systems}

\correspondingauthor{Billy Quarles}
\email{billylquarles@gmail.com}

\author[0000-0002-9644-8330]{Billy Quarles}
\affil{Center for Relativistic Astrophysics, School of Physics, 
Georgia Institute of Technology, Atlanta, GA 30332 USA}

\author[0000-0001-8308-0808]{Gongjie Li}
\affil{Center for Relativistic Astrophysics, School of Physics, 
Georgia Institute of Technology, Atlanta, GA 30332 USA}

\author[0000-0001-9786-1031]{Veselin Kostov}
\affil{NASA Goddard Space Flight Center, Greenbelt, MD 20771, USA}
\affil{SETI Institute, 189 Bernardo Avenue, Suite 200, Mountain View, CA 94043, USA}

\author[0000-0002-5234-6375]{Nader Haghighipour}
\affil{Institute for Astronomy, University of Hawaii-Manoa, Honolulu, HI 96822, USA}

\begin{abstract}
Planets that orbit only one of the stars in stellar binary systems (i.e., circumstellar) are dynamically constrained to a limited range of orbital parameters and thus understanding conditions on their stability is of great importance in exoplanet searches.  We perform $\sim$700 million N-body simulations to identify how stability regions depend on properties of the binary, as well as, the starting planetary inclination and mean longitude relative to the binary orbit.  Moreover, we provide grid interpolation maps and lookup tables for the community to use our results.  Through Monte-Carlo methods we determine that planets with a semimajor axis $a_p$ $\lesssim$8\% of the binary semimajor axis $a_{bin}$ will likely be stable given the known distribution of binary star parameters.  This estimate varies in the Lidov-Kozai regime or for retrograde orbits to 4\% or 10\% of $a_{bin}$, respectively.  Our method to quickly determine the circumstellar stability limit is important for interpreting observations of binaries using direct imaging with JWST, photometry with TESS, or even astrometry with Gaia.
\end{abstract}


\section{Introduction} \label{sec:intro}
Before the discovery and confirmation of the first exoplanet around a Sunlike star, 51 Peg b  \citep{Mayor1995,Marcy1995}, many theoretical investigations uncovered that planetary systems could stably orbit one star in a stellar binary despite the intense periodic forcing from the stellar companion \citep{Szebehely1980,Rabl1988,Benest1988a,Benest1988b,Benest1993}.  Around the same time, observers began using radial velocity techniques to probe for the existence of substellar companions and one of the first proposed candidates was $\gamma$ Cep Ab \citep{Campbell1988}.  The host star belonged to a binary system, $\gamma$ Cep AB, which has a binary separation of only about 20 AU.  \cite{Walker1992} later attributed the radial velocity signal to stellar rotation due to the limits of the data, but more observations eventually confirmed the existence of $\gamma$ Cep Ab \citep{Hatzes2003}.  Before this confirmation, four exoplanets were discovered (55 Cnc Ab, $\tau$ Boo Ab, $\upsilon$ And Ab, \& 16 Cyg Bb) whose host star is part of a more widely separated binary \citep{Butler1997,Cochran1997}.

The closest Sunlike stars to the solar system, $\alpha$ Cen AB, are part of a similar binary architecture to $\gamma$ Cep AB, where numerical studies have explored the extent by which a stable planetary orbit can persist around each star over a wide range of initial conditions \citep{Wiegert1997,AndradeInes2014,Quarles2016,Quarles2018a}.  However, the detection of planets in $\alpha$ Cen AB remains in dispute, where radial velocity observations of $\alpha$ Cen B suggested that an Earth-mass planet orbited the star on a $\sim$3.2 day orbit \citep{Dumusque2012}.  Later studies revealed that the significance of the radial velocity signal changed dramatically when different methods were used to analyze the data \citep{Hatzes2013,Rajpaul2016}.  The formation of Earth-mass planets around either star in $\alpha$ Cen AB is also contentious when considering the known planet population in binary systems\footnote{https://www.univie.ac.at/adg/schwarz/multiple.html} consists predominantly of Jupiter mass planets and the decades of radial velocity measurements of $\alpha$ Cen AB favor Neptune -- Saturn mass planets in the upper limit \citep{Zhao2018}.  Early models of planet formation in $\alpha$ Cen AB showed that solar system-like formation conditions (embryos \& planetesimals) could produce terrestrial planets \citep{Quintana2002,Haghighipour2007}.  In contrast, \cite{Thebault2008} and \cite{Thebault2009} performed simulations of planetesimal growth in the $\alpha$ Cen system, where they determined that eccentricity pumping from the binary companion largely prevented growth and subsequent planet formations processes would be very difficult.  Theoretical models by \cite{Zsom2011} also found that the presence of a binary companion would lower the total disk mass through truncation in addition to the problems of planetesimal growth.  However, it has been suggested that these issues could be avoided under a range of initial conditions including non-coplanar disks \citep{Marzari2009,Xie2010,Rafikov2013,Rafikov2015}.  Future observations with the James Webb Space Telescope (JWST) could help resolve these disputes with a detection or at least put stronger upper limits on the size of any potential worlds orbiting $\alpha$ Cen A \citep{Beichman2019}.

Ground based observational studies have indicated that Sunlike stars are common among binary systems, where nearly half of Sunlike stars have a binary companion \citep{Raghavan2010,Moe2017}.  The Kepler Space Telescope observed 2878 eclipsing binary systems, 1.3\% of all targets, within the prime mission \citep{Kirk2016} and discovered about a dozen \emph{circumbinary} planets.  Using the results from Kepler, \cite{Kraus2016} proposed the apparent lack of \emph{circumstellar} planets discovered with binary separations similar to $\gamma$ Cep AB was due to the ruinous effects of the binary star on the planet formation process.  However, \cite{Ngo2017} found no evidence that the presence or absence of stellar companions alters the distribution of planet properties when including radial velocity systems.  Observations from the redesigned Kepler mission, K2, uncovered at least two Neptune sized circumstellar planets (K2-136 \cite{Ciardi2018}; K2-288 \cite{Feinstein2019}) with projected binary separations $\sim$40 AU.  \cite{Martin2017} showed that orbital precession could be affecting the detectability of circumstellar planets through transit surveys and observations of many more binaries are needed to identify reliable population statistics.  Fortunately, the Transiting Exoplanet Survey Satellite (TESS \cite{Ricker2015}) is observing large portions of the sky and expected to observe $\sim$500,000 eclipsing binaries \citep{Sullivan2015}, thereby increasing the prospects of detecting circumstellar planets in binaries.

Most studies have used the stability limit formulas developed by \cite{Holman1999} to determine whether a circumstellar or circumbinary candidate within a binary system is \textit{bone fide} and not a false positive.  \cite{Holman1999} note the formulas have limitations; most notably that circular, coplanar test particles were used in its development and the expressions are valid to within $4-11\%$.  \cite{PilatLohinger2002} and \cite{PilatLhoinger2003} also performed a study for circumstellar planets using the Fast Lyapunov Indicator \citep{Froeschle1997} and explored a limited range of planets on inclined orbits.  {The stability of circumbinary orbits were investigated including the effects of mutual inclination \citep{Doolin2011} and planet packing \cite{Kratter2014,Quarles2018b}.  Those approaches have even been applied to the Pluto-Charon system \citep{Kenyon2019a,Kenyon2019b} allowing for better mass estimates of the satellites.}  New methods and updated formulas have substantially reduced the uncertainty in the stability limit for circumbinary planets (machine learning \cite{Lam2018}; grid-based interpolation \cite{Quarles2018b}). {We implement a similar method proposed by \citealt{Quarles2018b} to update the stability limit for circumstellar planets.}  

In this work, we extend the grid-based interpolation method to massive circumstellar planets, identify changes to the stability limit relative to the mutual inclination of the planet, and estimate the probability density function given the known distribution of binary stars along with a prospective critical semimajor axis ratio $a_c/a_{bin}$.  Our numerical setup is described in Section \ref{sec:method}, which includes our definition for stability and starting conditions.  In Section \ref{sec:stab_limit}, we determine a revised stability limit {considering circular and eccentric binaries and accounts for a significant mutual inclination between the planetary and binary orbital planes.  In addition, we investigate methods to utilize our results through interpolation maps and look-up tables.}  Section \ref{sec:max_ecc} explores more stringent conditions on the stability limit due to the secular forcing of eccentricity from the stellar binary.  Section \ref{sec:pops} applies our revised stability limit to binary star population statistics and a system recently observed by TESS, LTT 1445ABC.  Finally, our results and concluding remarks are summarized in Section \ref{sec:conc}.

\section{Methods}\label{sec:method}

\subsection{Defining Stability and the Stability Limit}
One of the defining features when expanding from the two body (Kepler) problem to the three body problem (TBP) is the emergence of chaos \citep[see details in][]{Mudryk2006,Musielak2014}, or a sensitivity to initial conditions where the future state of a system is indeterminate after a sufficient time.  The possible outcomes within the TBP are broadly defined as stable (i.e., the system is bounded in space and consists of three bodies for all time) or unstable (i.e., at least one body is no longer bounded in space or a collision occurs).  Notice that these definitions do not include chaos explicitly, where bundles of initial conditions can be either stable or unstable, as well as chaotic.  There is a transition between stable and unstable states that has been identified within astronomical systems \citep{Lecar1992,Smith1993,Gozdziewski2001,PilatLohinger2002,Cincotta2003,Cuntz2007,Eberle2008,Quarles2011,Giuppone2012,Satyal2013} using chaotic indicators similar to the Lyapunov exponent  \citep{Benettin1980,Gonczi1981,Froeschle1997,Cincotta1999,Cincotta2000}.

Due to this transition region, we must be more precise about our definitions for stability and allow for a more probabilistic consideration.  Moreover, we cannot compute the evolution of a system for all time and thus we define a given initial condition as potentially stable if the planet survives for a predefined timescale ($\sim10^5$ yr) around its host star.  {Previous works \citep{David2003,Fatuzzo2006,Quarles2016} showed that most of the initial conditions that will become unstable over longer timescales exist near mean motion resonances at the edge of stability and our approach excludes the resonant region due to its dependence on the initial mean longitude of the planet.}    Unstable initial conditions are those that do not survive for the required timescale due to a collision of the planet with either of the host stars or the distance to the host star exceeds 200 AU (i.e., twice the largest binary semimajor axis considered).   

Observations of planets in binaries are usually limited so that a full initial condition is not available. \cite{Rabl1988} and \cite{Holman1999} accounted for this by prescribing their stability criterion in terms of a set of observables that are the most readily obtained (e.g., binary mass ratio, binary eccentricity, semimajor axis ratio).  The binary mass ratio $\mu$ ($=M_B/(M_A + M_B)$) and the binary eccentricity $e_{bin}$ can be deduced from photometric and/or radial velocity observations.  The semimajor axis ratio $a_p/a_{bin}$ can be determined if the respective orbital periods are well determined, otherwise projected values are typically used.  Due to the long orbital periods involved and small number of observations, not many other observables are typically known.  As a result, \cite{Holman1999} assumed that the orbital planes of the binary and planet are aligned (i.e., $i_p = i_{bin}=\omega_p = \Omega_p = \omega_{bin} = \Omega_{bin} = 0^\circ$) and the planetary orbit is initially circular (i.e., $e_p = 0$).  Additionally, the restricted TBP was implemented, where a \emph{massless test particle} was used for the planet as a matter of numerical efficiency. 

A combination of parameters ($\mu$, $e_{bin},$ and $a_p/a_{bin}$) are numerically evolved for eight equally-spaced planetary phase angles and a critical semimajor axis $a_c$ is determined, if all eight of the trials are deemed stable.  Most of our simulations follow this general approach, except we use an Earth-mass particle that gravitationally interacts with the binary and evaluate 91 initial planetary phase angles so that we can investigate the probabilistic transition region between stable and unstable orbits.  \emph{The stability limit $a_c$ is defined in our analysis as the largest semimajor axis ratio $a_p/a_{bin}$, where all 91 initial phase angles ($0^\circ\leq \lambda_p \leq 180^\circ$) survive up to our predefined timescale.} 

\subsection{Setup for Numerical Simulations} \label{sec:numerical}
The numerical simulations in our study use the \texttt{mercury6} integration package that has been modified so that the orbits of planets in binaries are efficiently evaluated \citep{Chambers2002}.  This well-established coded was developed for planet formation simulations within $\alpha$ Centauri AB \citep{Quintana2002}, where a hybrid scheme allows for switching between a symplectic integration method (i.e., Wisdom-Holman splitting \citep{Wisdom1991,Wisdom1992}) and an adaptive method (e.g., Bulirsch-Stoer \citep{Press1992}).  We check a subset of our runs using a newer n-body code, \texttt{REBOUND}, with the IAS15 integrator \citep{Rein2012,Rein2015} to ensure the accuracy of our results.

We setup our simulations starting from a set of unitless parameters ($\mu$, $e_{bin},$ and $a_p/a_{bin}$), which allows for scalability of our results to many different dynamical systems.  {The range of these parameters are given in Table \ref{tab:init_cond} along with the range of inclination and mean longitudes for defining an orbit.} The total mass of the binary is equal to one Solar mass ($M_A+M_B=1$ M$_\odot$) and the individual stellar masses are determined via $\mu$ ($=M_B/(M_A + M_B)$).  The initial planetary semimajor axis $a_p$ is 1 AU and the initial value of the semimajor axis ratio is used to determine the appropriate binary semimajor axis ($\sim1-100$ AU).  The initial phase of the binary begins at periastron ($\lambda_{bin} = 0^\circ$), which lies on the positive x-axis ($\omega_{bin} = 0^\circ$) within the orbital plane of the binary.  All of our simulations use the binary orbital plane as a reference ($i_{bin}=\Omega_{bin}=0^\circ$), where the initial planetary inclination $i_p$ is the relative angle between the planetary and binary orbital planes.  Moreover, these orbital planes begin nodally aligned ($\Omega_p = \Omega_{bin}$).

\begin{deluxetable}{lcc}
\tablecaption{Range of Initial Conditions Used in our Simulations  \label{tab:init_cond}}
\tablehead{ \colhead{parameter} & \colhead{range} & \colhead{step}}
\startdata
$\mu$ & 0.01 -- 0.99 & 0.01\\
$e_{bin}$ & 0.00 -- 0.80 & 0.01\\
$a_p/a_{bin}$ & 0.010 -- 0.800 & 0.001\\
$i_p$ & $0^\circ - 180^\circ$ & $30^\circ$\\
$\lambda_p$ & $0^\circ - 180^\circ$ & $2^\circ$\\
\enddata
\tablecomments{The range of mass ratio $\mu$ also includes two additional extreme values, 0.001 and 0.999. The range of planetary inclination $i_p$ is restricted to $0^\circ$, $30^\circ$, $45^\circ$, and $180^\circ$ for our simulations with eccentric binaries. }
\end{deluxetable}

The accuracy of our simulations is controlled by choosing a timestep that is 5\% of the planetary orbital period $T_p$ ($\propto (1-\mu)^{-1/2}$).  Stability studies of $\alpha$ Cen AB \citep{Wiegert1997,Quarles2016} have shown that the region of stability does not change appreciably until $i_p$ is greater than 40$^\circ$ and the extent of stability increases for retrograde orbits.  There is an intermediate region ($40^\circ \gtrsim i_p \gtrsim 140^\circ$), where the Lidov-Kozai (LK) mechansim \citep{Lidov1962,Kozai1962} drives large eccentricity oscillations that can reduce stability zones unless the planet and binary are apsidally misaligned by 90$^\circ$ \citep{Giuppone2017}.  We probe a wide range of planetary inclinations (0$^\circ$, 30$^\circ$, 60$^\circ$, 85$^\circ$, 95$^\circ$, 120$^\circ$, 150$^\circ$, and 180$^\circ$) for circular binaries and then we limit our investigation of eccentric binaries to 4 planetary inclinations: 0$^\circ$, 30$^\circ$, 45$^\circ$, and 180$^\circ$.  The mass ratio $\mu$ is varied uniformly in steps of 0.01 from 0.01 -- 0.99, where we also evaluate the special cases of 0.001 and 0.999.  The binary eccentricity $e_{bin}$ is sampled in 0.01 steps from 0 -- 0.8, while the semimajor axis ratio $a_p/a_{bin}$ ranges from 0.01 -- 0.8 in steps of 0.001.  While the initial phase of the binary remains fixed ($\lambda_{bin}=0^\circ$), the mean longitude of the planet is varied in 2$^\circ$ steps from 0$^\circ$ -- 180$^\circ$.  


\subsection{Symmetries in the Parameter Space}
\begin{figure*}
    \centering
    \includegraphics[width=\linewidth]{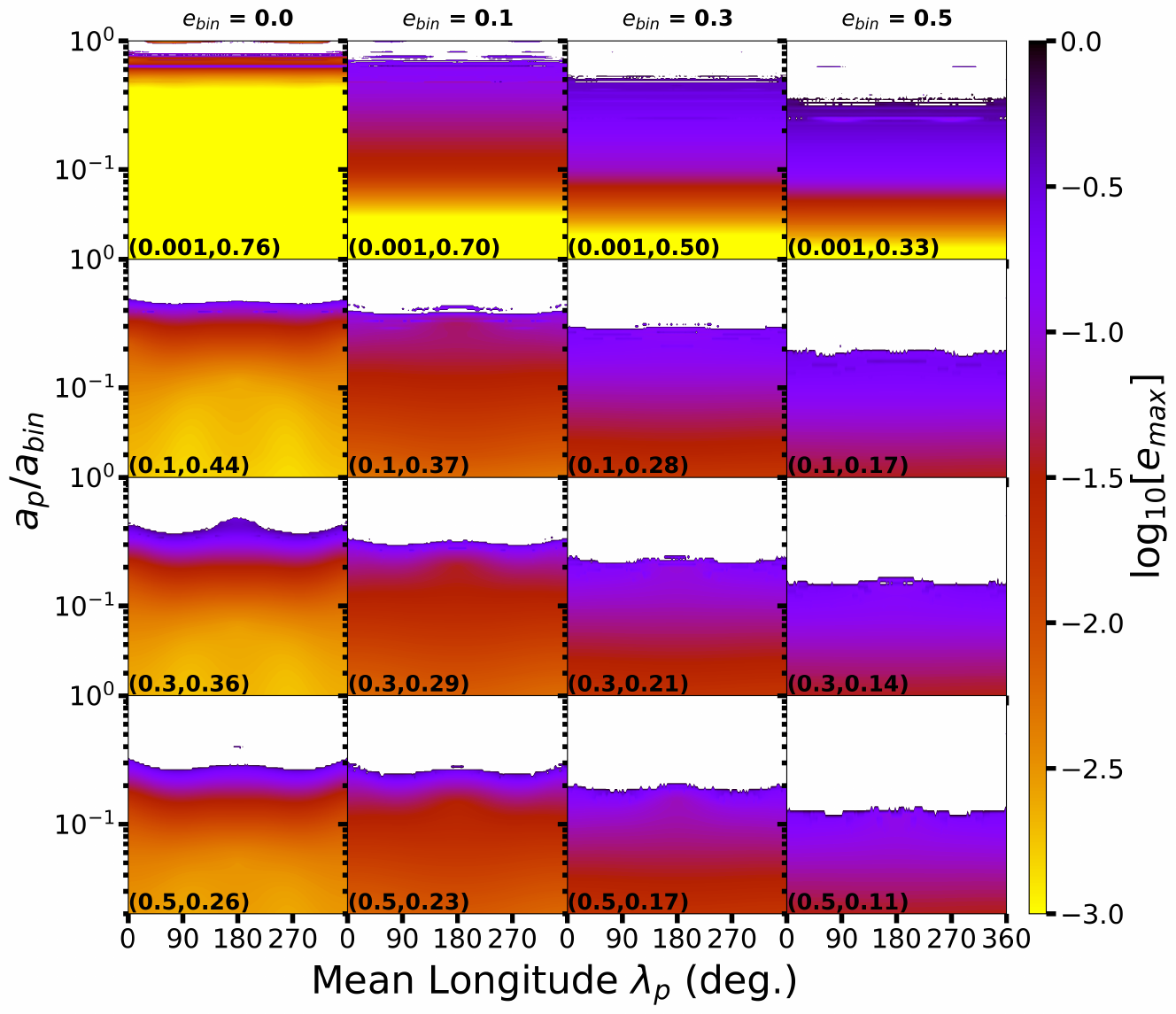}
    \caption{The maximum planetary eccentricity attained (color-coded on a logarithmic scale) for coplanar initial conditions that survive for $5\times10^5$ years of simulation time over a range of initial semimajor axis ratios $a_p/a_{bin}$ and mean longitudes $\lambda_p$.  {Each panel contains a label ($\mu$, $a_c$) indicating the host binary mass ratio and the critical semimajor axis ratio determined for each binary configuration, respectively.}  Note the mirror symmetry about $\lambda_p = 180^\circ$.  }
    \label{fig:Copl_stab}
\end{figure*}

The range of parameters that we explore is quite broad and is necessarily so, to provide the most general results.  Our previous study of circumbinary planets \citep{Quarles2018b} exploited a symmetry in the initial mean longitude, and allowed for the computations to be performed efficiently.  To evaluate whether the similar symmetries exist, we find the maximum eccentricity a planet attains when beginning from a circular orbit.  Figure \ref{fig:Copl_stab} illustrates the maximum eccentricity that a \emph{coplanar} planet attains for stable configurations as a function of the initial semimajor axis ratio $a_p/a_{bin}$ and mean longitude $\lambda_p$, where the white regions represent initial conditions that are unstable on a 500,000 yr timescale.  {The panels are labeled to show the mass ratio $\mu$ and stability limit $a_c$ determined.}  As the mass ratio increases (top to bottom), the stability limit $a_c$ decreases and the maximum eccentricity increases by an order of magnitude.  Additionally, the first column shows a symmetry in the mean longitude that appears more strongly with increasing $\mu$.  The growth of the maximum eccentricity is more pronounced when the binary eccentricity $e_{bin}$ increases (left to right).  Indeed, these trends are known from secular studies \citep[][and references therein]{AndradeInes2016,AndradeInes2017} and further indicates the accuracy of our numerical simulations.  These trends justify our numerical setup (see Section \ref{sec:numerical}) and particular restrictions in mean longitude when surveying the full parameter space.  

\begin{figure*}
    \centering
    \includegraphics[width=\linewidth]{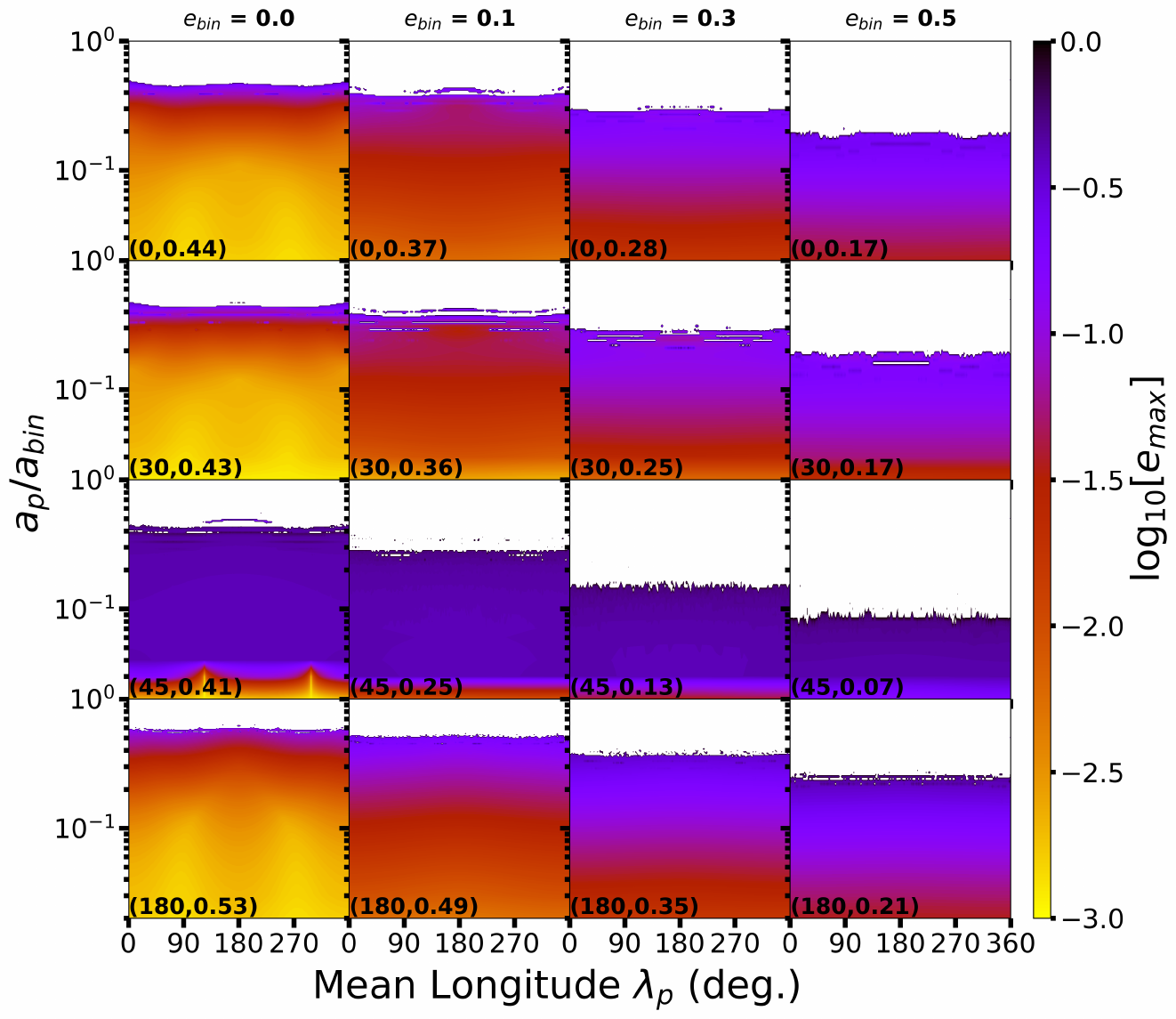}
    \caption{Similar to Figure \ref{fig:Copl_stab}, but the mass ratio is constant ($\mu=0.1$) and the planetary inclination is varied.  {Each panel contains a label ($i_p$, $a_{c}$) indicating the planetary inclination and the critical semimajor axis ratio, respectively.}  Note the mirror symmetry about $\lambda_p = 180^\circ$ remains for inclined orbits.  }
    \label{fig:Incl_stab}
\end{figure*}

Other dynamical effects, such as the Lidov-Kozai (LK) mechanism \citep{Lidov1962,Kozai1962}, can arise when considering inclined orbits and affect how much we can exploit particular symmetries.  In Figure \ref{fig:Incl_stab}, we perform similar simulations as in Fig. \ref{fig:Copl_stab}, but for inclined orbits with a constant mass ratio ($\mu=0.1$).  {The planetary inclination $i_p$ and determined stability limit $a_c$ are given as tuples in the lower left of each panel.}  The difference between a planet inclined by $30^\circ$ relative to the coplanar case is relatively minor, where variations occur mainly at large semimajor axis ratio that depend on the initial mean longitude.  When the planetary inclination is increased to $45^\circ$, the stability limit $a_c$ decreases more substantially relative to the coplanar case and the typical maximum eccentricity is $\sim$0.5 across most semimajor axis ratios.  Finally, our retrograde ($i_p=180^\circ$) runs demonstrate a $\sim$25\% increase in the stability limit $a_c$ relative to the respective coplanar runs.   This is roughly what one would expect from previous studies of the restricted TBP, where stable particles fill a larger portion of the Hill radius \citep{Henon1970}.  Overall, Figs. \ref{fig:Copl_stab} and \ref{fig:Incl_stab} demonstrate that the range of mean longitude required to determine the stability limit can be reduced to $0^\circ-180^\circ$ independent of the initial planetary inclination $i_p$ and possibly a separate condition can be placed on the maximum eccentricity, where $e_{max}\lesssim0.5$.

\section{Revised Stability Limit}\label{sec:stab_limit}
{To} meaningfully revise the previous determination of stability limit $a_c$ for circumstellar planets in binaries \citep{Holman1999}, we perform a large number of numerical simulations\footnote{The results of our simulations are publicly available on \dataset[zenodo.org]{http://doi.org/10.5281/zenodo.3579202} as a series of compressed tar archives.} ($\sim$700 million in total) that report the lifetime of a planet and the maximum eccentricity attained for each set of initial conditions.  We use four input parameters ($\mu$, $e_{bin}$, $a_p/a_{bin}$, $\lambda_p$) that relate to the most easily obtained observables and iterate this process over a range of planetary inclinations (see Section \ref{sec:numerical}) that are representative of the dynamical interactions affecting the stability limit.  In this section, our analysis is divided into 3 parts: the case of circular binaries ($e_{bin}=0)$, comparing to previous results by \cite{Holman1999} with elliptical binaries ($e_{bin}\leq 0.8$), and the interpolation maps across each planetary inclination.

\subsection{Circular Binaries} \label{sec:circ_stab}
One of the oldest problems studied within orbital dynamics is the circular restricted three body problem (CRTBP), where the two more massive primaries begin on circular orbits and the third body is not massive enough to substantially alter the orbit of the primaries (see \cite{Musielak2014} for a review).  Due to the restriction on the binary orbit's eccentricity, an integral arises that reduces some of the complexity.  The solution to the integral, or Jacobi constant $C_J$ in the CRTBP, provides a natural means to orient our analysis.  The so-called zero velocity contour (ZVC) for a given mass ratio $\mu$ and semimajor axis ratio $a_p/a_{bin}$, describes a topological boundary that limits the trajectory of a planet and can be related to the Jacobi constant \citep{Eberle2008}.

\begin{figure*}
    \centering
    \includegraphics[width=0.58\linewidth]{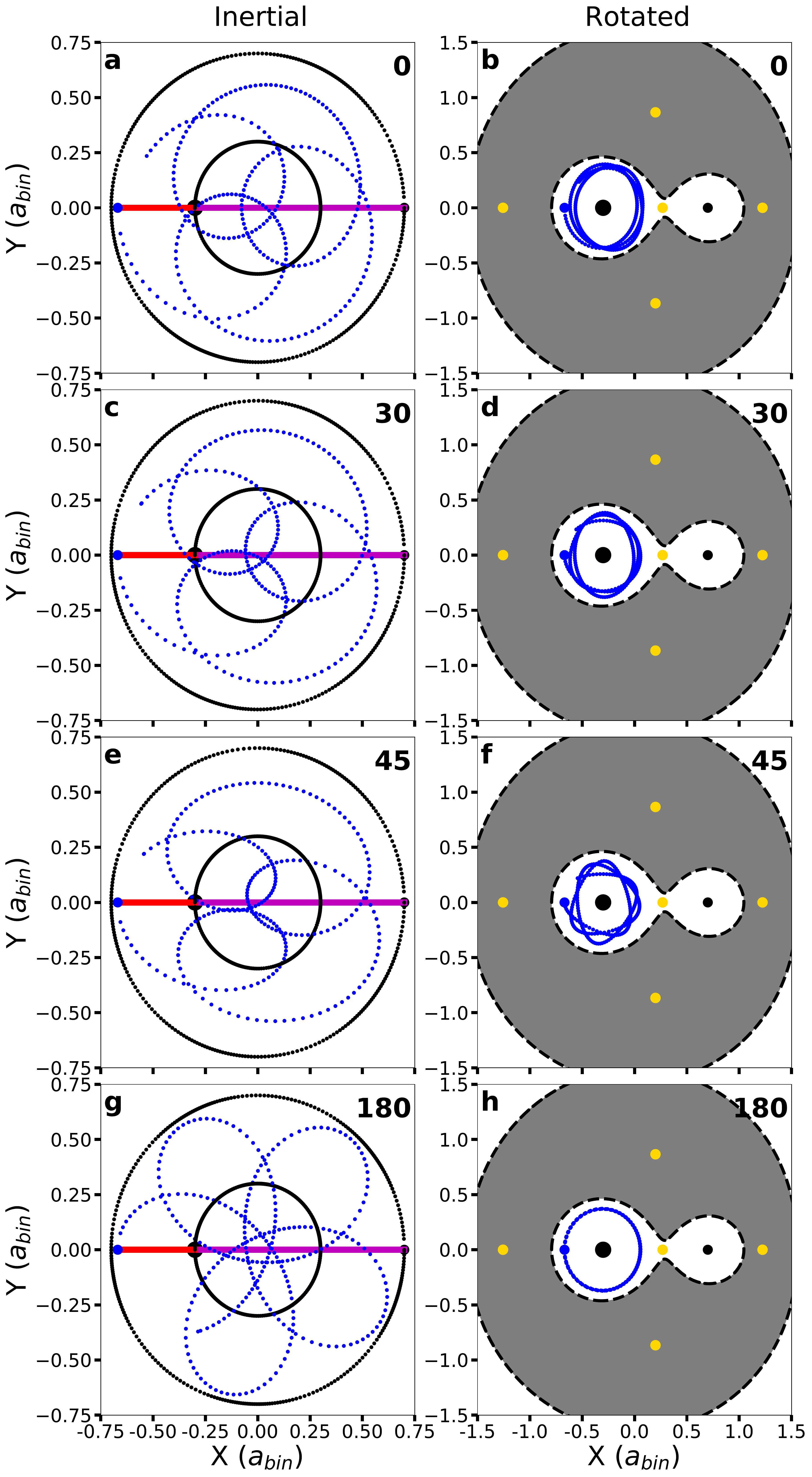}
    \caption{The trajectories in the X-Y plane for a three body system ($\mu = 0.3$, $a_p/a_{bin} = 0.369$, and $\lambda_p = 180^\circ$) in inertial (left) and rotated (right) coordinates{ for a range of mutual inclinations in degrees (top right)}.  The initial distance of the planet (blue) to the host star is marked by a red line, where the binary semimajor axis is indicated by a magenta line.  The trajectory of the planet over one orbit of the more massive primary is given in each coordinate system, where the planet begins inclined to the X-Y plane as indicated in the top right corner.  The gold points in the right column correspond to the Lagrange points.  The black dashed contours denote the boundaries of zero velocity in the rotated coordinates and the forbidden region is shaded gray. {Note the difference in scale between the two columns.}}
    \label{fig:ZVC_incl}
\end{figure*}

We use this formalism to explore the changes in the trajectory for a single initial condition ($\mu = 0.3$, $a_p/a_{bin} = 0.369$, and $\lambda_p = 180^\circ$) with respect to a changing planetary inclination.  Figure \ref{fig:ZVC_incl} shows each  trajectory over 1 orbit of the binary using inertial, or static, coordinates (left column) and rotated coordinates (right column), where the rotation rate is equal to the binary mean motion.  The nominal location of the $N:1$  mean motion resonances (MMRs), in terms of semimajor axis ratio, is $a_p/a_{bin} = ((1-\mu)/N^2)^{1/3}$, which places the initial condition between the 4:1 and 3:1 MMRs.  For the coplanar case in Fig. \ref{fig:ZVC_incl}a, we trace the trajectory of the planet to nearly complete 3 orbits within 1 binary orbit.  This results in a quasi-periodicity for the trajectory in the rotated coordinates (Fig. \ref{fig:ZVC_incl}b) and is bounded by the ZVC.  When the planet is inclined by $30^\circ$ in Fig. \ref{fig:ZVC_incl}c, it follows a similar trajectory within a tilted plane and small differences appear due to the projection onto the X-Y plane.  The planetary orbit is more eccentric (not just in appearance due to projection) once the inclination is increased to 45$^\circ$ in Fig. \ref{fig:ZVC_incl}e.  A retrograde ($i_p = 180^\circ$) planet executes the most regular trajectory, where almost 5 orbits are now completed within a single binary orbit (Fig. \ref{fig:ZVC_incl}g) and the trajectory is periodic in rotated coordinates (Fig. \ref{fig:ZVC_incl}h).  From these example trajectories, we expect that the stability results for planetary inclinations up to $30^\circ$ to not differ too much from those determine for a coplanar orbit.  The increased eccentricity from the Lidov-Kozai mechanism will introduce a dependence on the initial mean longitude due to apsidal precession and retrograde orbits will be stable to significantly larger semimajor axis ratios.

\begin{figure*}
    \centering
    \includegraphics[width=\linewidth]{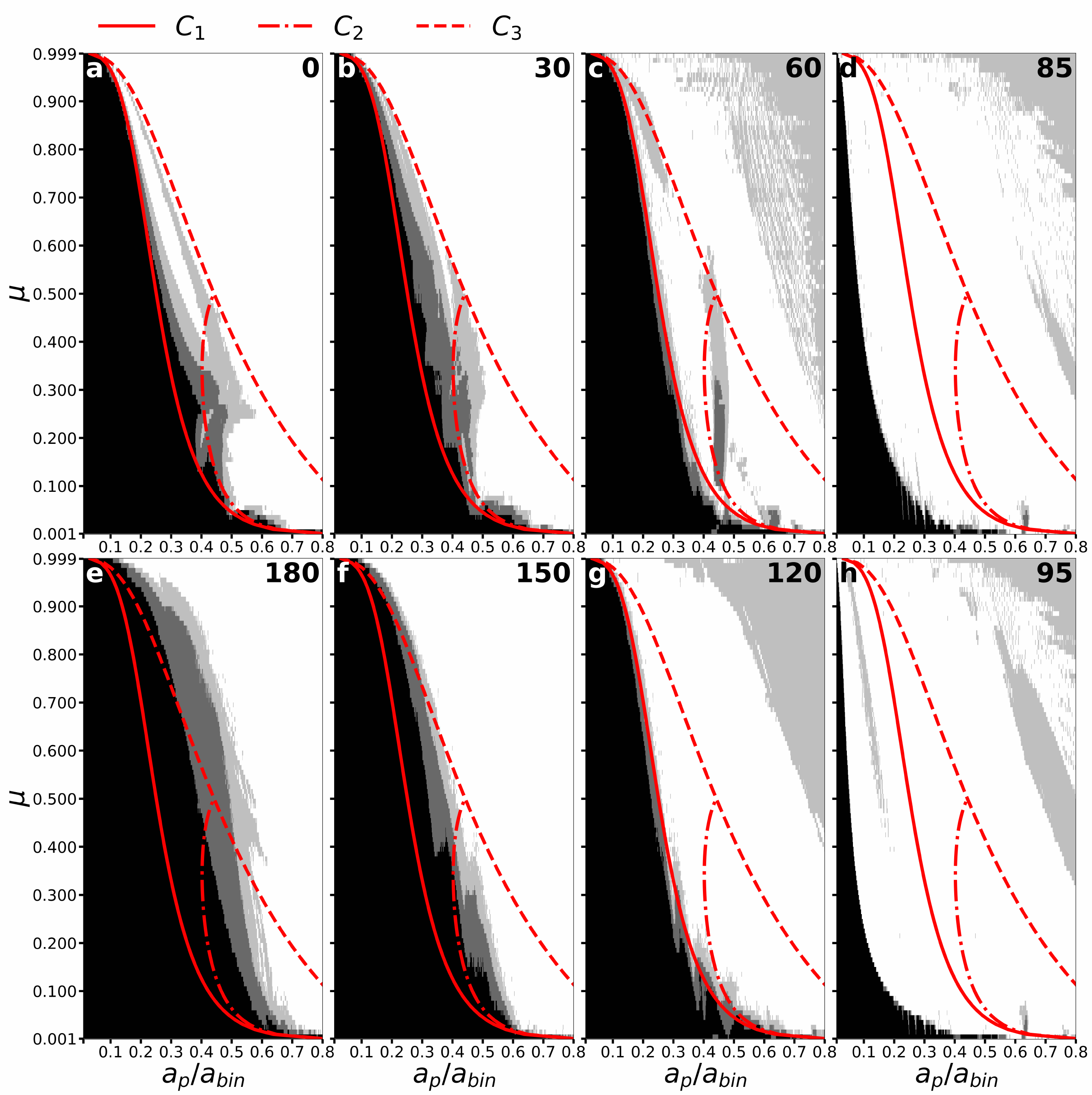}
    \caption{The stability regions within the CRTBP for a range of initial values in the mass ratio $\mu$ and semimajor axis ratio $a_p/a_{bin}$ considering a potentially inclined planet with the respective inclination in the top right corner.  The black cells indicate that all 91 trials survive for the entire 500,000 yr of the simulation (i.e., stable), where the dark or light gray cells mark those when either 20--99\% or 1--19\% of the trials survive, respectively. The white cells denote initial conditions, where none of trials survive (i.e., unstable).  The red curves correspond to openings in the zero velocity curve at each Lagrange point \citep{Eberle2008}. }
    \label{fig:CRTBP}
\end{figure*}

In the CRTBP some regions of the parameter space are more chaotic, which appear once the ZVC opens at the Lagrange point $L_1$ which lies between the stars \citep{Quarles2011}.  Additional chaos can arise when the other collinear Lagrange points no longer lie within the forbidden region due to changes in the mass ratio $\mu$ or the semimajor axis ratio $a_p/a_{bin}$.  The origin of chaos in the CRTBP lies in the overlap of the $N:1$ MMRs with the binary, once the planetary eccentricity is substantially excited by a resonance \citep{Murray1999,Mardling2008}.  We take a wide-view of the possible initial conditions ($\mu$, $a_p/a_{bin}$, and $\lambda_p$) to produce probabilistic maps for a range of  planetary inclinations in Figure \ref{fig:CRTBP}.  The probability for stability is calculated simply as the percentage of simulations (out of 91 values for $\lambda_p$) that survive for 500,000 yr for a given mass ratio and semimajor axis ratio.  As a result, we identify regions that are stable (black), unstable (white), and quasi-stable (light or dark gray).  The quasi-stable region is split into two regimes where either 1--18 trials survive (1--20\%; light gray) or 19--90 trials survive (20--99\%; dark gray) to delineate between chaotic zones and possible resonant islands that strongly depend on $\lambda_p$.  In Fig. \ref{fig:CRTBP}, each panel also marks the critical boundaries (red curves) for which the ZVC opens for a collinear Lagrange point ($L_1$, $L_2$, or $L_3$) as detailed in \citep{Eberle2008} using the respective Jacobi constant ($C_1$, $C_2$, or $C_3$).

\cite{Eberle2008} and \cite{Quarles2011} explored the coplanar case (Fig. \ref{fig:CRTBP}a) in detail using the Jacobi constant and maximum Lyapunov component, respectively, for planets that orbit the more massive primary ($\mu \leq 0.5$).  However, Fig. \ref{fig:CRTBP}a shows that using the Jacobi constant $C_1$ as a stability criterion agrees well for all $\mu$ and only a few stable (black) regions exist to the right of the solid red curve, $C_1$.  Between the curves for $C_1$ and $C_3$, there are initial conditions that can be stable depending on the initial mean longitude due to MMRs \citep{Morais2012,Morais2013}, as indicated by the quasi-stable (light and dark gray) region and \cite{Quarles2011} identified this regime as chaotic using the maximum Lyapunov exponent.  All of the initial values to the right of $C_3$ (dashed red curve) are unstable due to large excitations of the planetary eccentricity by the perturbing star and the opening of all the collinear Lagrange points in the ZVC. 

After the planet is inclined 30$^\circ$ relative to the binary orbit (Fig. \ref{fig:CRTBP}b), the stable (black) region does not change much.  Figs. \ref{fig:CRTBP}c and \ref{fig:CRTBP}d illustrate significant differences from the coplanar cases due to the LK mechanism, where a large periodic eccentricity oscillation occurs that erodes much of the parameter space between the $C_1$ and $C_3$ curves.  The stable regime for Fig. \ref{fig:CRTBP}d lies significantly below the $C_1$ curve demonstrating the effectiveness of the LK mechanism.  Some resonant regions (i.e., N:1 MMRs) can remain stable for longer periods, but must maintain a special configuration (e.g., similar to Pluto's 3:2 MMR with Neptune).  

Figs. \ref{fig:CRTBP}e--\ref{fig:CRTBP}h show how the stability regions vary for retrograde orbits ($i_p^{retro} = 180^\circ - i_p$), where the stable retrograde initial conditions (black) in Fig. \ref{fig:CRTBP}e represent a much larger fraction of the parameter space and do not agree with the Jacobi constant stability criterion from \cite{Eberle2008}.  The extent of the quasi-stable regime (light or dark gray) also increases, but it reaches a limit at the 3:2 MMR with the binary orbit.  Fig. \ref{fig:CRTBP}f can be largely constrained with the Jacobi constant and appears very similar in extent to Fig. \ref{fig:CRTBP}b.  Figs. \ref{fig:CRTBP}g and \ref{fig:CRTBP}h are similar to their prograde counterparts (Figs. \ref{fig:CRTBP}c and \ref{fig:CRTBP}d), but the stabilizing effects from starting in retrograde increases the extent of the quasi-stable region at high $\mu$ and $a_p/a_{bin}$.

\begin{figure}
    \centering
    \includegraphics[width=\linewidth]{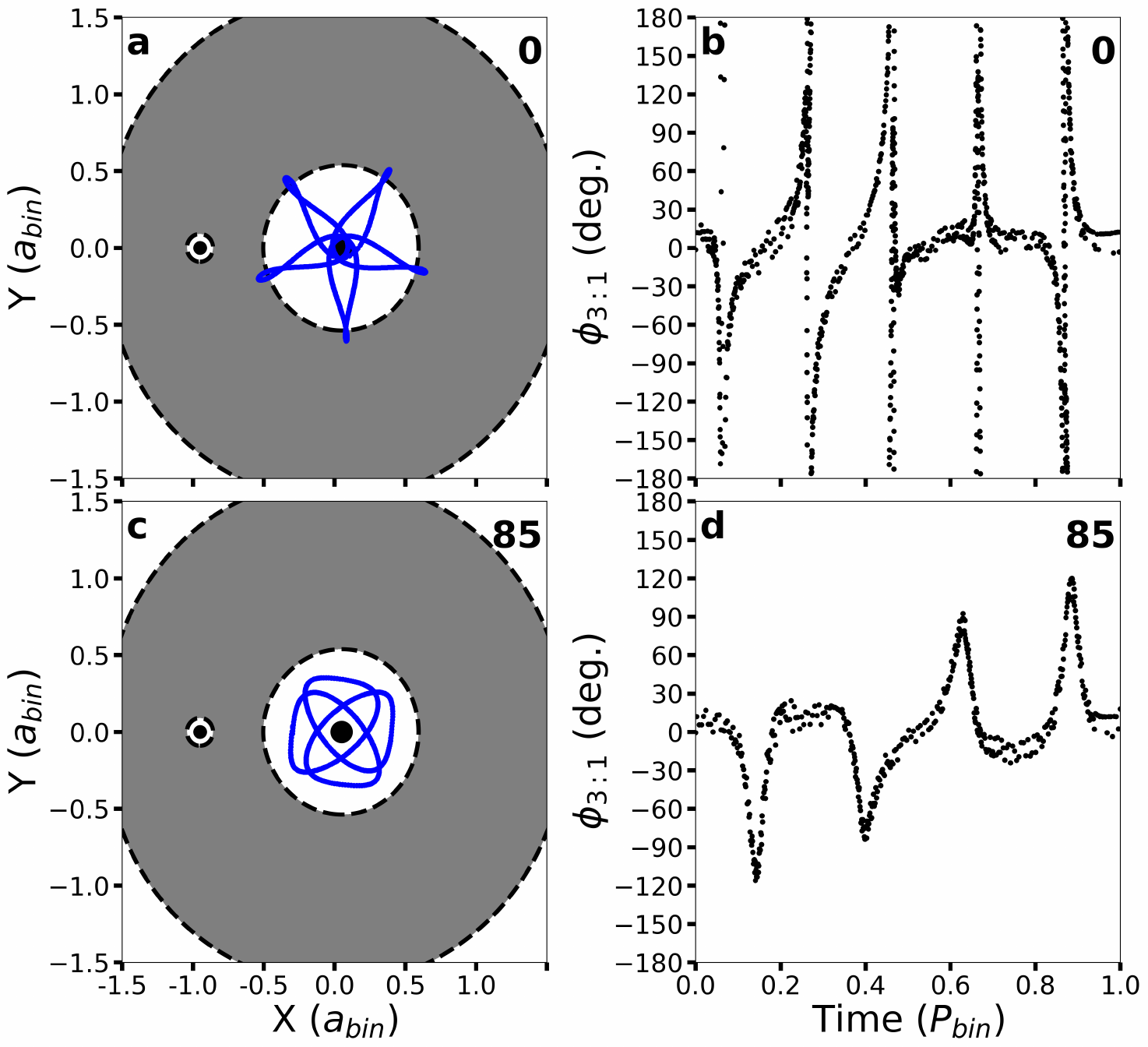}
    \caption{The trajectory of a coplanar or highly inclined planet in rotated coordinates over a single orbit of the binary beginning with $a_p/a_{bin} = 0.78$, a host binary with $\mu = 0.95$, and an initial mean anomaly $MA=33^\circ$. Panel (a) differs from panel (c) in the starting planetary inclination (indicated in upper right), which manifests in a difference in the strength of the Coriolis force.  The starting condition lies within the Hill sphere of the more massive star (near origin) and can result in the possible capture in the 3:1 MMR.  Panels (b) and (d) illustrate the evolution of the resonance angle for the 3:1 MMR, $\phi_{3:1} = 3\lambda_{bin} - \lambda_p - 2\varpi_{bin}$. }
    \label{fig:CRTBP_LK}
\end{figure}

Another special circumstance can arise when the planet begins with a significant orbital velocity out of the binary plane so that the Coriolis force is minimized.  In this scenario, the initial orbital velocity is calculated assuming the lower mass star is the host (i.e., high $\mu)$, but the high semimajor axis ratio $a_p/a_{bin}$ starts the planet within the Hill sphere of the more massive star.  As a result, the planet can be captured into N:1 MMRs by the primary.  Although this setup is allowed (i.e., no physical laws broken), it could be very unlikely to occur within known pathways for planet formation and we highlight it for completeness.  Figure \ref{fig:CRTBP_LK} illustrates the evolution of a particular initial condition in rotated coordinates that can lead to this type of capture where the planet begins either coplanar (Fig. \ref{fig:CRTBP_LK}a) or highly inclined (Fig. \ref{fig:CRTBP_LK}c).  To confirm a capture into the 3:1 MMR, we plot the evolution of the resonance angle $\phi_{3:1}$, where there are points in the coplanar case (Fig. \ref{fig:CRTBP_LK}b) that are ill-defined (i.e., vertical jumps) and the highly inclined simulation (Fig. \ref{fig:CRTBP_LK}d) shows libration.

Even in the simplified case of the CRTBP, there is not a simple expression to define the likelihood that a planetary orbit will be stable.  The most straight-forward prescription to estimate stability is N-body simulations on a case-by-case basis, but this process can be greatly expedited or avoided using our results.  For a nearly \emph{coplanar} planet within a circular binary (e.g., transiting an eclipsing binary), the stability limit can be estimated using the Jacobi constant at the first Lagrange point $C_1$.  We provide an algorithm to determine this curve as follows:
\begin{enumerate}
    \item Choose a value of $\mu$ from a list.
    \item Approximate through a power series or numerically determine the Lagrange point $L_1$ using a given $\mu$ and a root finding function.
    \item Use the location of $L_1$ to determine the Jacobi constant $C_1$ using the equation for the zero velocity surface ($2U = C_J$; see Chap. 3 of \cite{Murray1999})
    \item Use the value of $C_1$ from (3) and numerically solve Equation 14 from \cite{Eberle2008} using a root finding function.
    \item Go back to (1), until the end of the list is reached.
\end{enumerate}

\noindent Table \ref{tab:circ_stab} lists values for the semimajor axis ratio and mass ratio where openings of the ZVC occur at $L_1$ or $L_3$ using the above algorithm and these values correspond to the red curves ($C_1$ or $C_3$, respectively) in Fig. \ref{fig:CRTBP}.  For a moderately  \emph{inclined} planet within a binary, similar curves provide a good approximation for the stability limit but a coplanar retrograde orbit can significantly expand the stability limit as shown by $C_1^{retro}$.  Finally, we provide python tools on GitHub to query the data set represented in Fig. \ref{fig:CRTBP} and there is a routine that returns the probability for stability through grid interpolation for a given combination of $\mu$ and $a_p/a_{bin}$.  

\begin{deluxetable}{lccc}
\tablecaption{Stability Limits in the CRTBP using the Mass ratio {and Jacobi Constant Criterion}  \label{tab:circ_stab}}
\tablehead{ \colhead{$\mu$} & \colhead{$C_1$} & \colhead{$C_3$} & \colhead{$C_1^{retro}$}}
\startdata
0.001	&	0.7988	&	0.9978	& 0.780\\
0.01    &   0.6368  &   0.9785  & 0.689\\
0.05	&	0.4906	&	0.9012	& 0.582\\
0.10	&	0.4230	&	0.8201	& 0.549\\
0.15	&	0.3825	&	0.7515	& 0.519\\
0.20	&	0.3533	&	0.6921	& 0.496\\
0.25	&	0.3301	&	0.6399	& 0.475\\
0.30	&	0.3107	&	0.5932	& 0.455\\
0.35	&	0.2937	&	0.5509	& 0.438\\
0.40	&	0.2784	&	0.5120	& 0.422\\
0.45	&	0.2644	&	0.4760	& 0.405\\
0.50	&	0.2511	&	0.4421	& 0.389\\
0.55	&	0.2385	&	0.4100	& 0.373\\
0.60	&	0.2261	&	0.3792	& 0.357\\
0.65	&	0.2137	&	0.3492	& 0.340\\
0.70	&	0.2011	&	0.3195	& 0.322\\
0.75	&	0.1880	&	0.2898	& 0.302\\
0.80	&	0.1738	&	0.2592	& 0.281\\
0.85	&	0.1579	&	0.2266	& 0.256\\
0.90	&	0.1387	&	0.1899	& 0.225\\
0.95	&	0.1119	&	0.1436	& 0.180\\
0.99    &   0.0682  &   0.0790  & 0.107\\
0.999	&	0.0329	&	0.0353	& 0.048\\
\enddata
\tablecomments{{Semimajor axis ratio $a_p/a_{bin}$ of the stability limit for a given mass ratio $\mu$ and using the Jacobi constant $C_1$ and $C_3$ as a stability criterion at the Lagrange points $L_1$ and $L_3$, respectively}.  $C_1^{retro}$ marks the numerically determined stability boundary for retrograde orbits ($i_p = 180^\circ$) due to a limitation of the Jacobi constant criterion{, where the uncertainties in these values are 0.001}. }
\end{deluxetable}

\subsection{Eccentric Binaries}\label{sec:ecc_stab}
Planets orbiting a single star of a circular binary is only a subset of all the possibilities and thus, we extend our method to investigate binary systems with a significant eccentricity ($e_{bin} \leq 0.8$).  This extra dimension expands the volume of our parameter space by almost 2 orders of magnitude and thus we perform our numerical simulations with computational efficiency in mind.  {In Section \ref{sec:circ_stab}, we explore 8 different mutual inclinations because the eccentricity forcing on the planet is mainly due to MMRs.  For eccentric binaries, we limit our investigation to only 4 mutual inclinations ($i_p = 30^\circ$, $45^\circ$, and $180^\circ$) as the LK mechanism greatly reduces the stability limit within the $40^\circ - 140^\circ$ range of inclination \citep{Quarles2016,Quarles2018a}.}  

Our coplanar simulations ($i_p=0^\circ$) use the uniform stepping described in Section \ref{sec:numerical} to determine the stability boundary $a_c$.  To make the exploration of the inclined cased more efficient, we use the method of \cite{Lam2018}, where they focused on a range in semimajor axis ratio (i.e., window) that would most likely contain the stability limit.  To do this, we use the results from the coplanar runs to determine an appropriate trial window for the other inclined cases ($i_p = 30^\circ$, $45^\circ$, and $180^\circ$).  We then numerically investigate for longer timescales (500,000 yr) using semimajor axis ratios within  $0.5a_c^0-1.5a_c^0$, where $a_c^0$ is the stability limit in the respective prograde, coplanar simulation.  If all the simulations in this range $0.5a_c^0-1.5a_c^0$ are unstable, then we explore a range of smaller semimajor axis ratios ($0.001 \leq a_p/a_{bin} \leq 0.5a_c^0$).  This procedure was sufficient to cover regions of parameter space that we expect to either be truncated (e.g., Lidov-Kozai mechanism for $i_p = 45^\circ$) or expanded (e.g., retrograde orbits for $i_p = 180^\circ$).

\begin{figure}
    \centering
    \includegraphics[width=\linewidth]{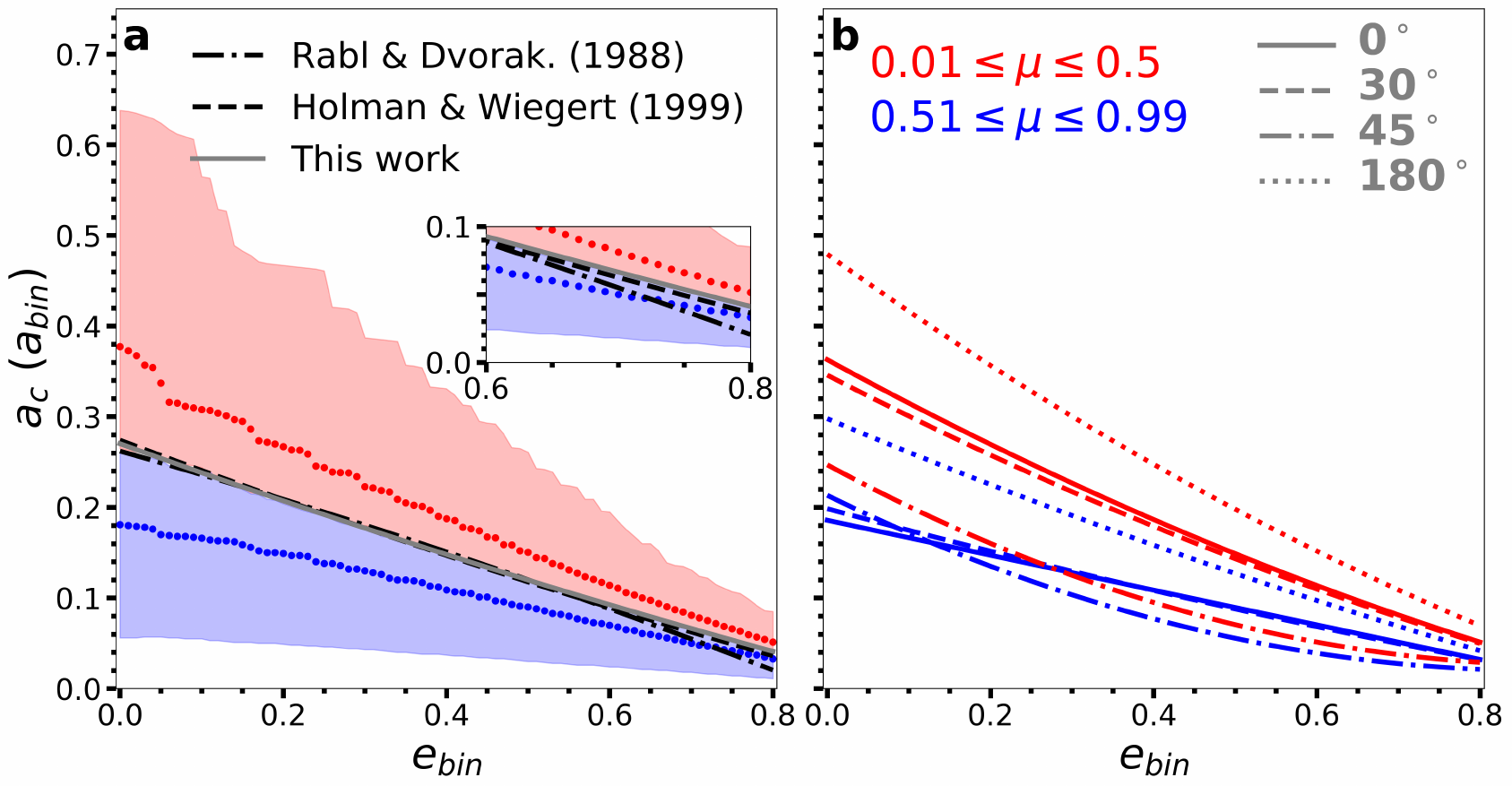}
    \caption{Critical semimajor axis $a_c$ as a function of the binary eccentricity $e_{bin}$ for a planet orbiting either: the more (red) or less (blue) massive star.  The black curves in panel (a) illustrate the previous results for $\mu = 0.5$, where our results are given by the solid gray curve.  The inset shows a magnified view for high binary eccentricity ($0.6\leq e_{bin} \leq 0.8$), where the black and gray curves no longer overlap.  The median value for $a_c$ (color-coded points) are shown in panel (a), where the color-coded area signifies the full range with respect to each domain.  The curves in panel (b) demonstrate the least squares best-fit for the median $a_c$ as a function of each planetary inclination. }
    \label{fig:mu_approx}
\end{figure}

Following \cite{Holman1999}, we take these results and compare to previous studies \citep{Rabl1988} that prescribe a stability limit for equal mass stars ($\mu = 0.5$).  Figure \ref{fig:mu_approx} illustrates how our results compare for the special case of equal mass stars with a coplanar planet, the range of stability for 2 regimes of $\mu$ (red and blue), and the changes to the median stability limit within each regime for each planetary inclination.  \cite{Rabl1988} performed simulations for only 300 binary periods and explored up to $e_{bin} = 0.6$.  Thus, we expect for deviations to appear for high binary eccentricities ($0.6 \leq e_{bin} \leq 0.8$ where the expression from \cite{Rabl1988} is extrapolated as seen in Fig. \ref{fig:mu_approx}a.  There is also a small difference between the curve from \cite{Holman1999} (black dashed) and our results (solid gray) in the high binary eccentricity region, as well as in the $e_{bin}\leq 0.1$ region.  These differences are due to the factor of 10 increase in sampling that we perform and amount to slight changes in the stability limit for a coplanar planet.  The shaded regions (red and blue) in Fig. \ref{fig:mu_approx}a demonstrate the extent that the stability limit can vary as a function of $\mu$, where $\mu = 0.01$ follows the upper red boundary and $\mu = 0.99$ is much flatter along the lower blue boundary.  As in the case for the CRTBP, we can see that the stability limit is not symmetric about the equal mass case.

The planetary inclination also plays a role in determining the stability limit $a_c$, where Fig. \ref{fig:mu_approx}b shows these differences through quadratic fits to the median stability limit for each $e_{bin}$.  The $i_p = 0^\circ$ (solid) and $i_p = 30^\circ$ (dashed) curves are nearly identical in Fig. \ref{fig:mu_approx}b and there is not a strong dependence on $\mu$ when split into the two regimes (red and blue).  As a result, the stability limit for coplanar planets will be similar enough, in most cases, to be valid up to $\sim40^\circ$, until the Lidov-Kozai mechanism becomes active.  The same will be true for retrograde planetary inclinations ($140^\circ \leq i_p < 180^\circ$) relative the exactly retrograde case ($i_p = 180^\circ$).  Sufficiently inclined planets ($i_p = 45^\circ$, dash-dot in Fig. \ref{fig:mu_approx}b), where the Lidov-Kozai mechanism is active have a lower median stability limit due to eccentricity excitation and it is largely independent of $\mu$.  On the other hand, planets in retrograde orbits (dotted in Fig. \ref{fig:mu_approx}b) have a larger median stability limit.  

Table \ref{tab:ecc_stab} provides coefficients and uncertainties for the quadratic fitting formula ($a_c/a_{bin} = c_1 + c_2 e_{bin} + c_3 e_{bin}^2$) from \cite{Rabl1988}, \cite{Holman1999}, and using the stability limits determined through our simulations.  The difference between our coplanar results ($i_p = 0^\circ$) and the previous studies is that we break the fitting formula into 2 regimes for $\mu$ as motivated by the CRTBP and Fig. \ref{fig:mu_approx}.  If we find the average value for each coefficient ($c_1 - c_3$), then our results are consistent with the previous works.  Our $c_1$ and $c_2$ coefficients for $i_p=30^\circ$ are similar to our coplanar coefficients.  The coefficients for the $i_p = 45^\circ$ fitting formula depend more strongly on the binary eccentricity due to the Lidov-Kozai mechanism.  The coefficients for the retrograde fitting formulas are larger than the coplanar, prograde formulas by more than 30\%.  Moreover, each set of coefficients reveal that the stability limit could be up to two times bigger for planets that orbit the more massive primary over the less massive secondary star. 

\begin{deluxetable}{lcccc}
\tablecaption{Critical Semimajor Axis as a Function of the Binary Eccentricity $e_{bin}$  \label{tab:ecc_stab}}
\tablehead{\colhead{} & \colhead{$i_p$} & \colhead{$c_1 \pm \sigma_1$} & \colhead{$c_2 \pm \sigma_2$} & \colhead{$c_3 \pm \sigma_3$}}
\startdata
$0.1\leq\mu \leq 0.9^\dagger$ & $0^\circ$ & $0.262 \pm 0.006$& $-0.254 \pm 0.017$ &  $0.060 \pm 0.027$  \\
$0.1\leq\mu \leq 0.9^\star$ & $0^\circ$ & $0.274 \pm0.008$ & $-0.338 \pm 0.045$ &  $0.051 \pm 0.055$ \\
\hline
$0.01\leq \mu\leq 0.5$ & $0^\circ$ & $0.363 \pm 0.001$ & $-0.492\pm0.006$ &  $0.129 \pm 0.005$  \\
$0.51 \leq \mu \leq 0.99$ & $0^\circ$ & $0.186 \pm 0.001$ & $-0.193 \pm 0.003$ &  $0.001\pm 0.003$ \\
\hline
$0.01\leq \mu\leq 0.5$ & $30^\circ$ & $0.346 \pm 0.002$ & $-0.464 \pm 0.006$ &  $0.117 \pm 0.005$  \\
$0.51 \leq \mu \leq 0.99$ & $30^\circ$ & $0.198 \pm 0.001$ & $-0.243 \pm 0.002$ &  $0.043 \pm 0.002$ \\
\hline
$0.01\leq \mu\leq 0.5$ & $45^\circ$ & $0.247 \pm 0.005$ & $-0.487 \pm 0.018$ &  $0.268 \pm 0.016$  \\
$0.51 \leq \mu \leq 0.99$ & $45^\circ$ & $0.213 \pm 0.002$ & $-0.441 \pm 0.009$ &  $0.252 \pm 0.008$ \\
\hline
$0.01\leq \mu\leq 0.5$ & $180^\circ$ & $0.479 \pm 0.002$ & $-0.647 \pm 0.008$ &  $0.168 \pm 0.008$  \\
$0.51 \leq \mu \leq 0.99$ & $180^\circ$ & $0.298 \pm 0.001$ & $-0.378 \pm 0.006$ &  $0.072 \pm 0.006$ \\
\enddata
\tablecomments{{The coefficients ($c_1-c_3$ and uncertainties($\sigma_1-\sigma_3$)} from $^\dagger$\cite{Rabl1988} and $^\star$\cite{Holman1999} are listed that use a quadratic fitting function, $a_c/a_{bin} = c_1 + c_2 e_{bin} + c_3 e_{bin}^2$.  We use the same function in this work, but we split it between 2 domains in $\mu$ for each planetary inclination $i_p$.}
\end{deluxetable}

An accurate and reliable fitting formula should include a dependence on the mass ratio $\mu$ given the asymmetries we have described in the CRTBP and in Fig. \ref{fig:mu_approx}.  \cite{Holman1999} assumed that the stability limit depends linearly on $\mu$ and quadratically for $e_{bin}$.  We follow the same prescription and provide updated values for the coefficients $c_1 - c_6$ using our much larger dataset in Table \ref{tab:form_stab} along with the previous values from \cite{Holman1999} for comparison.  Moreover, coefficients are listed for inclined planets ($i_p = 30^\circ$, $45^\circ$ and $180^\circ$), where we see that the coefficients for the coplanar and $30^\circ$ formulas are similar.  When the planet is initially inclined to $45^\circ$, the $c_3 - c_6$ terms dominate due to the strong dependence on $e_{bin}$ from the Lidov-Kozai mechanism.  The $c_1$, $c_3$, and $c_5$ coefficients for retrograde are larger than the respective coefficients for $i_p = 0^\circ$ formula, which indicates that retrograde stability depends more strongly on $e_{bin}$ due to a lower magnitude forcing on the planet from the stellar companion \citep{Henon1970}.

\begin{deluxetable}{lccccccc}
\tablecaption{Coefficients for the Critical Semimajor Axis  \label{tab:form_stab}}
\tablehead{\colhead{} & \colhead{$i_p$} & \colhead{$c_1 \pm \sigma_1$} & \colhead{$c_2  \pm \sigma_2$} & \colhead{$c_3 \pm \sigma_3$} & \colhead{$c_4 \pm \sigma_4$} & \colhead{$c_5 \pm \sigma_5$} & \colhead{$c_6 \pm \sigma_6$}}
\startdata
Previous Work$^\star$ & $0^\circ$ & $0.464\pm0.006$ & $-0.380\pm0.010$ &  $-0.631\pm0.034$ &  $0.586\pm0.061$ &  $0.650\pm0.041$ &  $-0.198\pm0.074$ \\
This work & $0^\circ$ & $0.501\pm0.002$ & $-0.435\pm0.003$ &  $-0.668\pm0.009$ &  $0.644\pm0.015$ &  $0.152\pm0.011$ &  $-0.196\pm0.019$ \\
This work & $30^\circ$ & $0.485\pm0.002$ & $-0.405\pm0.003$ &  $-0.684\pm0.009$ &  $0.603\pm0.015$ &  $0.190\pm0.011$ &  $-0.182\pm0.019$ \\
This work & $45^\circ$ & $0.428\pm0.002$ & $-0.318\pm0.003$ &  $-1.128\pm0.011$ &  $0.987\pm0.020$ &  $0.839\pm0.014$ &  $-0.825\pm0.024$ \\
This work & $180^\circ$ & $0.617\pm0.001$ & $-0.457\pm0.002$ &  $-0.787\pm0.008$ &  $0.586\pm0.014$ &  $0.163\pm0.010$ &  $-0.128\pm0.017$ \\
\enddata
\tablecomments{{The coefficients ($c_1 -c_6$) and uncertainties ($\sigma_1 -\sigma_6$)} from $^\star$\cite{Holman1999} are listed using the fitting formula, $a_c/a_{bin} = c_1 + c_2 \mu + c_3 e_{bin} + c_4 \mu e_{bin} + c_5 e_{bin}^2 + c_6 \mu e_{bin}^2$.  We use the same function in this work, but also fit for a range of planetary inclination $i_p$.}
\end{deluxetable}

\subsection{Interpolation Maps and Look-up Tables}
The fitting formulas in Section \ref{sec:ecc_stab} are dependent on the quality and breadth of the numerical simulations that underlie their derivation.  They are reliable in characterizing a population of exoplanets, but they can also be inaccurate when describing particular systems.  For instance, we can compare the case when $e_{bin} = 0$ in Table \ref{tab:form_stab} to the results in Table \ref{tab:circ_stab}.  The estimate for the critical semimajor axis $a_c$ from \cite{Holman1999} and our own newly updated estimation is different by more than 10\% for $\mu < 0.05$.  As a result, we suggest a different approach using a look-up table or interpolation map that relies less on statistical averaging which occurs in deriving a single fitting formula.  

\begin{figure*}
    \centering
    \includegraphics[width=\linewidth]{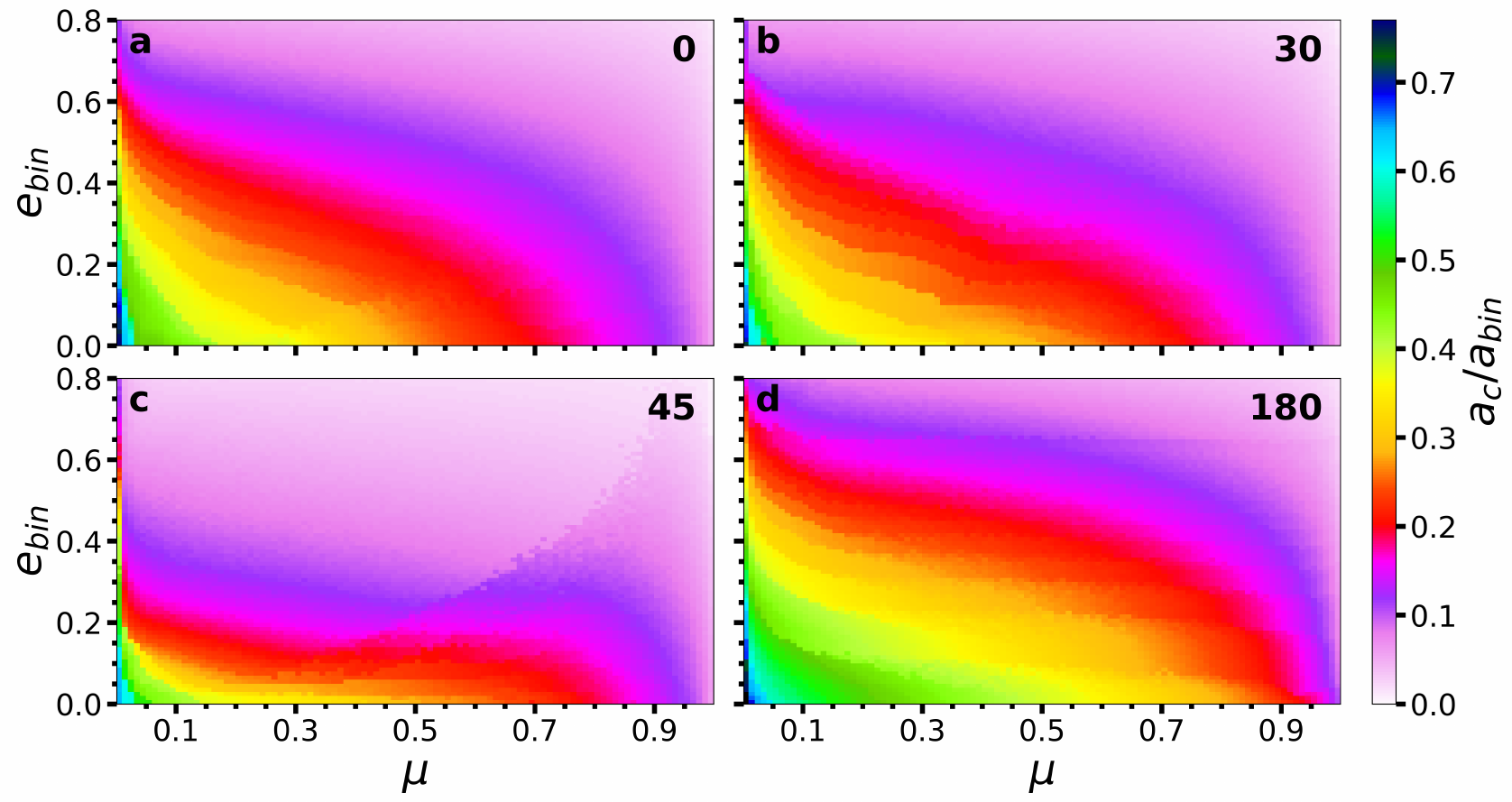}
    \caption{Critical semimajor axis $a_c$ (color-coded) as a function of the binary eccentricity $e_{bin}$ and the mass ratio $\mu$.  The initial planetary inclination $i_p$ is denoted in each panel in the upper right. }
    \label{fig:interp_maps}
\end{figure*}

Figure \ref{fig:interp_maps}\footnote{The data for these maps are available on \dataset[GitHub]{https://github.com/saturnaxis/ThreeBody_Stability}, along with an example python script that interpolates between grid values.} shows the critical semimajor axis ratio $a_c/a_{bin}$ (color-coded) over the full range of mass ratio $\mu$ and binary eccentricity $e_{bin}$.  Table \ref{tab:form_stab} shows only small differences in the fitting formulas for $i_p = 0^\circ$ and $i_p = 30^\circ$, where this trend is visually apparent in Figs. \ref{fig:interp_maps}a and \ref{fig:interp_maps}b.  In addition, Figs. \ref{fig:interp_maps}a and \ref{fig:interp_maps}b illustrate the asymmetry of the stability limit with respect to the mass ratio $\mu$, where for $\mu > 0.5$ the typical value is  $a_c/a_{bin} \lesssim 0.2$.  As a result, we find the volume for a stable planet for the primary star to host a nearly planar ($i_p \leq 30^\circ$) planet in a low mass ratio ($\mu < 0.3$) binary to be typically two times larger than for the lower mass secondary ($\mu > 0.7$).  The propensity for stable orbits is substantially reduced for a more inclined ($i_p = 45^\circ$) planet when $e_{bin} \gtrsim 0.2$ and the critical semimajor axis becomes more symmetric about $\mu = 0.5$ (Fig. \ref{fig:interp_maps}c).  A retrograde planet (Fig. \ref{fig:interp_maps}d) can be stable to much larger semimajor axis ratios and for a larger fraction of binary parameters ($\mu$ and $e_{bin}$){, which has been shown to some degree in several other investigations \citep{Henon1970,Benest1988b,Wiegert1997,Quarles2016}.}

 \section{More Stringent Stability Criteria}\label{sec:max_ecc}
 A more stringent stability {criterion} could be devised that depends on more parameters.  Part of the motivation for \cite{Holman1999} to use the binary mass ratio $\mu$ and eccentricity $e_{bin}$ was that these parameters are relatively easy to determine through an adequate number of observations through radial velocity or photometric surveys.  However, the Kepler era has uncovered 1000s of exoplanets \citep{Coughlin2016} and eclipsing binary stars \citep{Kirk2016}.  As a result, it becomes reasonable to identify how the stability limit changes with respect to the planetary eccentricity $e_p$.  Statistical studies of the radial velocity planets before Kepler suggested that the mean planet eccentricity was approximately $0.3$ \citep{Shen2008}.  Subsequent studies using both the radial velocity planets and those discovered by the Kepler Space Telescope showed that this is indeed correct for `single' planetary systems and a lower mean eccentricity ($<e_p> \approx 0.05$) is appropriate for systems with multiple planets \citep{Xie2016,VanEylen2019}.  Moreover, the presence of binary companions does not affect this estimate \citep{VanEylen2019}.  
 
\begin{figure*}
    \centering
    \includegraphics[width=\linewidth]{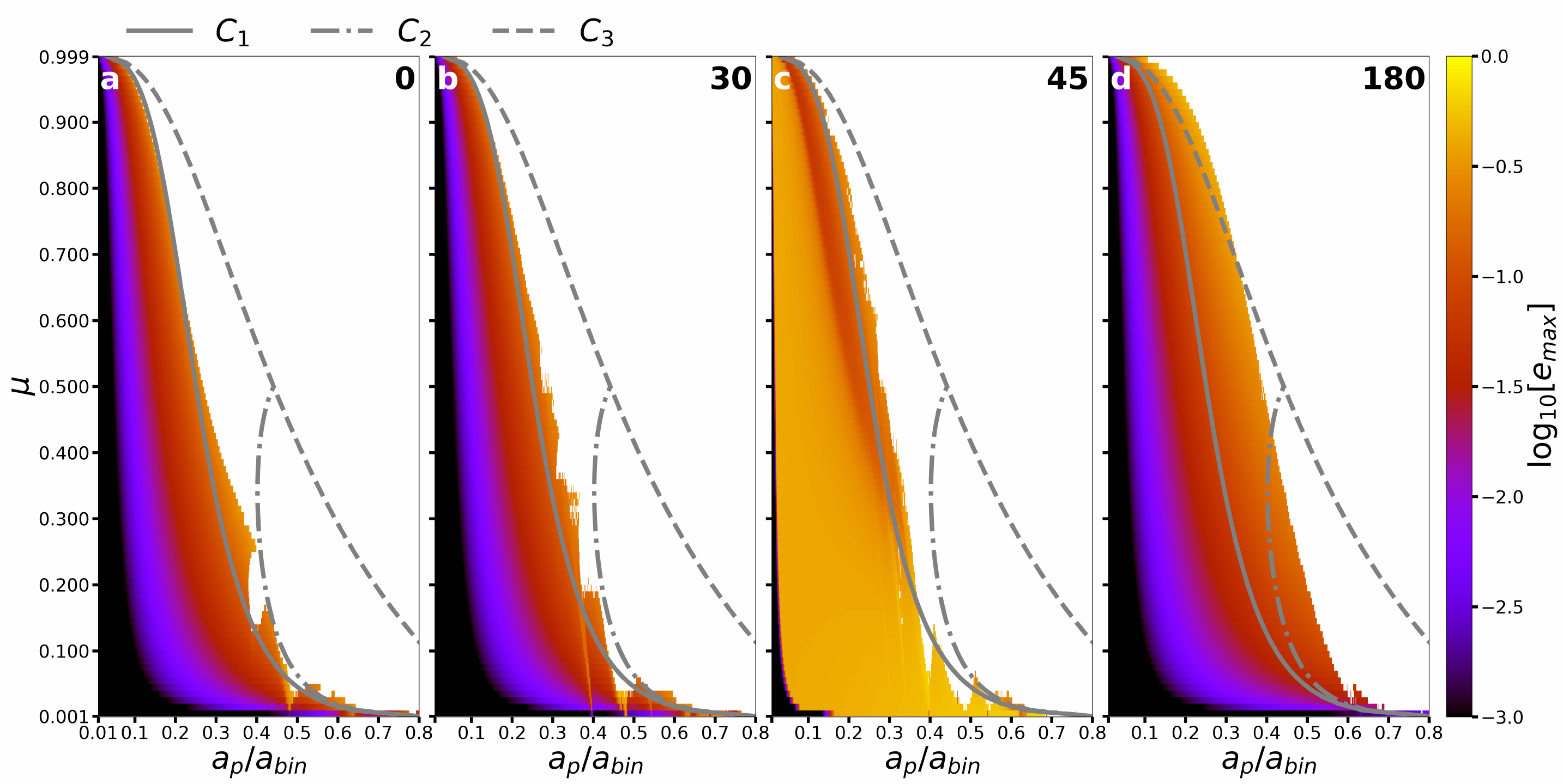}
    \caption{Similar to Figure \ref{fig:CRTBP}, but the stable points are color-coded (on a logarithmic scale) using the maximum planetary eccentricity attained over all 91 trials of mean anomaly.  The gray curves mark the opening of the zero velocity curve for each of the collinear Lagrange points.  }
    \label{fig:CRTBP_ecc}
\end{figure*}

 In the CRTBP, the maximum planet eccentricity is largely shaped by either a mean motion resonance or the Lidov-Kozai mechanism.  Both of these pathways for eccentricity excitation depend on the semimajor axis ratio $a_p/a_{bin}$ and less so on the mass ratio $\mu$.  Figure \ref{fig:CRTBP_ecc} illustrates the maximum planetary eccentricity $e_{max}$ attained (color-coded on a logarithmic scale) for the cases where the planetary stability is achieved for all 91 trials of mean longitude (i.e., black cells in Fig. \ref{fig:CRTBP}).  The gray curves in Fig. \ref{fig:CRTBP_ecc} mark the limits set by the Jacobi Constants at the collinear Lagrange points.  Figs. \ref{fig:CRTBP_ecc}a, \ref{fig:CRTBP_ecc}b, and \ref{fig:CRTBP_ecc}d show that the typical maximum planetary eccentricity is small ($e_{max} \lesssim 0.1$), where Fig. \ref{fig:CRTBP_ecc}c is large due to the Lidov-Kozai mechanism.  A substantial $e_{max}$ persists when $i_p = 45^\circ$ (Fig. \ref{fig:CRTBP_ecc}c) even for small semimajor axis ratios and mass ratios.
The CRTBP is a special case and nearly circular binaries are not likely to represent a large fraction of the general population of binary stars.  Detailed studies from secular theory \citep{Brouwer1961,Kaula1962,Heppenheimer1978,Murray1999,AndradeInes2016,AndradeInes2017} and N-body simulations \citep{Quarles2018a} have illustrated that the secular forcing from the stellar component induces a forced component to the planet's eccentricity that depends on the mass ratio $\mu$, semimajor axis ratio $a_p/a_{bin}$, and the binary eccentricity $e_{bin}$.  The planetary eccentricity varies about the forced eccentricity depending on the relative alignment ($\varpi_p - \varpi_{bin}$) with the binary orbit.  Each cell in Fig. \ref{fig:interp_maps}a represents a series of 91 trials with respect to the mean longitude and the $a_c/a_{bin}$ value determined is one where all trial mean longitudes survive for the full integration for a given $a_p/a_{bin}$.  We augment this stability criterion to include another condition that the maximum planetary eccentricity $e_{max}$ does not exceed 0.3 (i.e., the mean planetary eccentricity from observations) and signify this altered stability criterion as ($a_c/a_{bin})^\dagger$.

\begin{figure*}
    \centering
    \includegraphics[width=\linewidth]{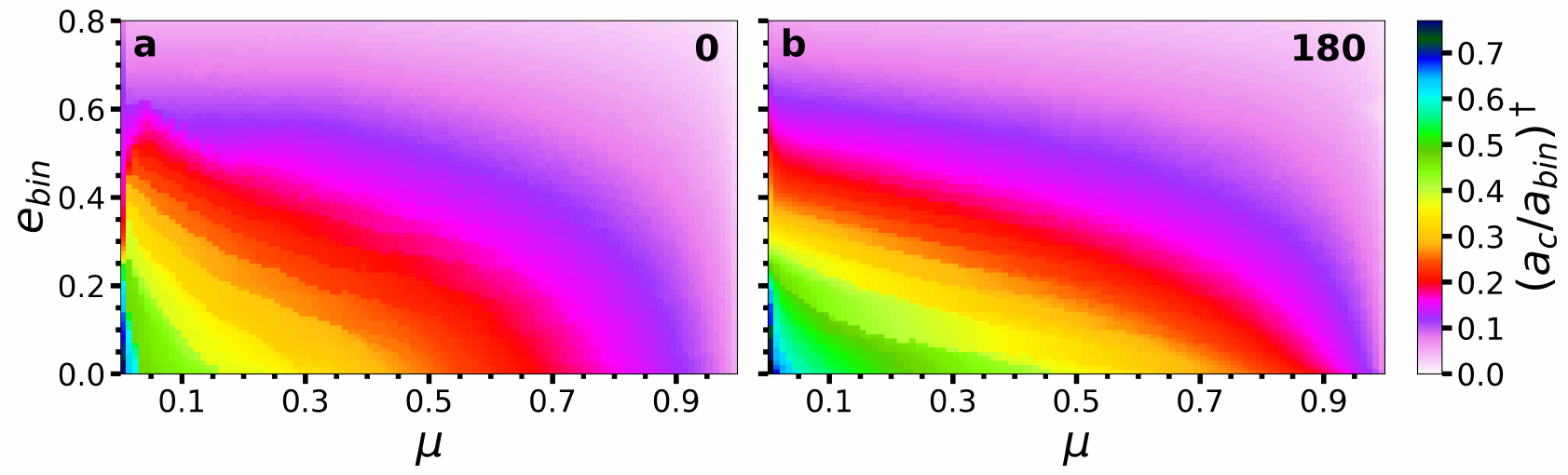}
    \caption{Similar to Figure \ref{fig:interp_maps}, but limited to a coplanar planet in (a) prograde or (b) retrograde relative to the binary orbit.  The color-code represents the critical semimajor axis $(a_c/a_{bin})^\dagger$, where the maximum planetary eccentricity is less than 0.3.  }
    \label{fig:SType_ecc}
\end{figure*}

The critical semimajor axis does not vary much when comparing Fig. \ref{fig:interp_maps}a to Fig. \ref{fig:interp_maps}b and the maximum planetary eccentricity will follow the same trend.  The maximum planetary eccentricity can be determined analytically, for nearly circular binaries, when the Lidov-Kozai mechanism is active \citep{Innanen1997} for $i_p=45^\circ$ and in most cases the maximum eccentricity is much larger than 0.3.  Therefore we only consider the coplanar configuration in either a prograde ($i_p=0^\circ$) or retrograde ($i_p=180^\circ$) direction relative to the binary orbit in Figure \ref{fig:SType_ecc}. Applying the extra condition on the planetary eccentricity typically decreases the critical semimajor axis by $\sim$10\% when comparing Fig. \ref{fig:interp_maps}a to Fig. \ref{fig:SType_ecc}a, where the decrease is much more substantial for the comparison between Fig. \ref{fig:interp_maps}d to \ref{fig:SType_ecc}b.  The differences between Fig. \ref{fig:SType_ecc}a and \ref{fig:SType_ecc}b are less dramatic than those in Fig. \ref{fig:interp_maps}, but planets orbiting within binaries with low eccentricity still retain their stability advantage because the forced component of eccentricity is also low.

The inclusion of an additional constraint like maximum eccentricity on planetary stability in binaries  decreases the critical semimajor axis ratio, where long period planets would be most affected.  However, observations of small planets have shown to come in multiples \citep{Fabrycky2014,Winn2015}, which suggests that planet multiplicity is a better additional constraint on the potential stability of a planet within a binary \citep{Quarles2018c} as the mean planet eccentricity of the observed exoplanets is lower \citep{Xie2016,VanEylen2019}.

\section{Binary Star Populations \& TESS}\label{sec:pops}
\subsection{Population Studies}
The results in Fig. \ref{fig:interp_maps} illustrates the variations of the critical semimajor axis with respect to the mass ratio $\mu$ and eccentricity $e_{bin}$, but observations of binary star populations, especially with Sunlike primaries, are not uniform with respect to these binary parameters \citep{Raghavan2010,Moe2017}.  Therefore, a statistical study is needed to gain further insights from our results using Monte Carlo approaches.  \cite{David2003} and \cite{Fatuzzo2006} performed a statistical study using the binary periastron distance as a stability criterion and used  the contemporary stability formula \citep{Holman1999} and a survey of binary stars \citep{Duquennoy1991}.  We follow their approach, {but use our our determination of the critical semimajor axis and more up-to-date results from binary star surveys \citep{Raghavan2010,Moe2017}}.  This is important for observations because when planets are discovered orbiting a star with a distant stellar companion, the binary period and eccentricity are often unknown (e.g., K2-288b \citep{Feinstein2019}), but the planetary semimajor axis and/or the projected binary semimajor axis can be determined (see Table \ref{tab:binaries} for a list of known planet hosting binary stars).  Moreover, observations of young binary stars can reveal the stability limit through dust and gas emission around either star \citep[e.g.,][]{Alves2019}, but this is also a projected distance.  

Our approach is to determine a probability distribution function (PDF) for the stability of planets in binaries as a function of the critical semimajor axis ratio $a_c/a_{bin}$, where the sum of the binary masses are equal to a Solar mass (i.e., $M_A + M_B = 1$ M$_\odot$).   The binary period (and semimajor axis) follows a log-normal probability distribution ($p_{\log P} \propto e^{-(\log P - \zeta)^2/2\sigma^2}$), where $\zeta = 5.03$ and $\sigma=2.28$ \citep{Raghavan2010}.  This probability distribution in \cite{Raghavan2010} has a broad range ($-2 \leq \log P \leq 10$, in days), where we limit our analysis to a smaller range in binary period ($4 \leq \log P \leq 7$) that corresponds to a reasonable range in binary semimajor axis (10 AU $\leq a_{bin} \leq$ 1000 AU).

Observers typically use the binary mass quotient $q$ ($=M_B/M_A$, where $M_B\leq M_A$), which algebraically relates to the dynamical mass ratio $\mu$ ($=q/(1+q)$).  From the mass quotient $q$, the range for $\mu$ is limited to 0.5 and our exploration of planets orbiting the secondary star ($\mu > 0.5$) is accomplished by subtracting the $\mu$ samples from unity (i.e., $1-\mu$).  We use a probability distribution ($p_q \propto q^\gamma$) for $q$ derived by \cite{Moe2017} that is a broken power law, which has $\gamma_1 = 0.3 \pm 0.4$ for $0.1\leq q \leq 0.3$ and $\gamma_2 = -0.5 \pm 0.3$ for $0.3 \leq q \leq 1$.  An excess twin fraction $F_{twin} = 0.1 \pm 0.3$ is added to the PDF for $q \geq 0.95$ (see their Fig. 2 \cite{Moe2017}).  A probability distribution ($p_e \propto e^\eta_{bin}$) for the binary eccentricity follows a single power law, which has $\eta = 0.4 \pm 0.3$ \citep{Moe2017}.  We note that the exponents used in these power laws are dependent on the range of sampled binary period $\log P$ \citep{Moe2017} and adjustments to the exponents are necessary if one expands our analysis to a broader range in binary period.

 \begin{figure*}
    \centering
    \includegraphics[width=\linewidth]{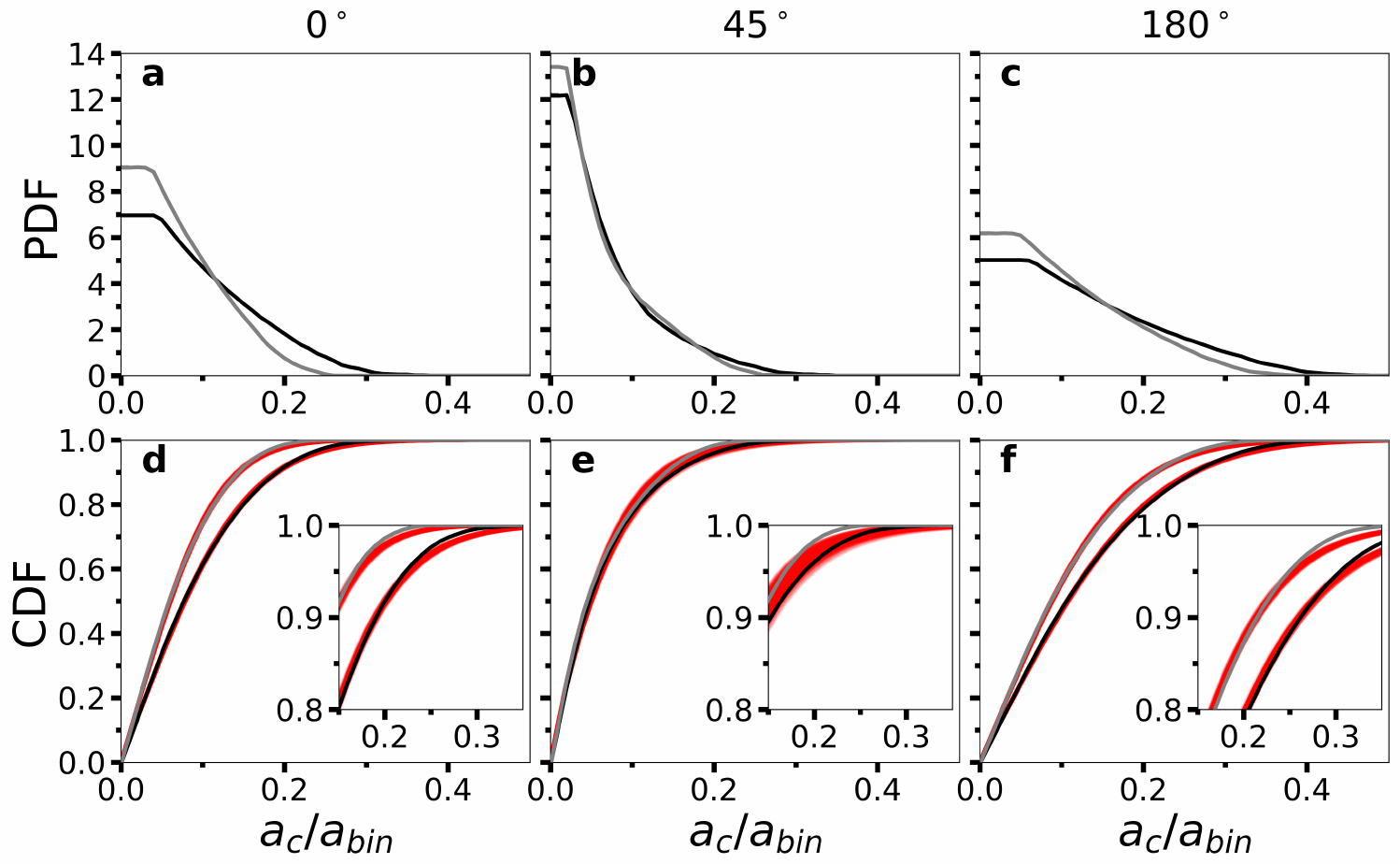}
    \caption{Probability Density Function (PDF) for the critical semimajor axis ratio $a_c/a_{bin}$ of a planet orbiting either Star A (black) or Star B (gray).  The PDF relative to whether the planet orbit is prograde ($i_p = 0^\circ$, a), undergoing Lidov-Kozai cycles ($i_p=45^\circ$, b), or is retrograde ($i_p=180^\circ$, c) with respect to the binary.  Panels (d)--(f) show the cumulative distribution function (CDF) for panels (a)--(c), respectively, and the zoomed insets show the top quintile of the CDF.  Samples from a numerical fit of the CDF using a modified exponential distribution (see main text) are marked in red, where the coefficients for the fit are given in Table \ref{tab:CDF_coeff}.  }
    \label{fig:SType_prob}
\end{figure*}

Using these probability distributions, we employ Monte Carlo integration to determine the fraction of observed binaries with a critical semimajor axis larger than a given $a_c/a_{bin}$ using the data in Figs. \ref{fig:interp_maps}a, \ref{fig:interp_maps}c, and \ref{fig:interp_maps}d.  We do not use Fig. \ref{fig:interp_maps}b because of its strong similarity to Fig. \ref{fig:interp_maps}a.  Fig. \ref{fig:interp_maps}a illustrates that all binaries within the domain will have a critical semimajor axis larger than 0.001 and very few binaries will have a $a_c/a_{bin} > 0.5$ where this is independent of the weighting given by the probability distributions we choose.  As a result, we integrate $a_c/a_{bin} =$ 0.001 -- 0.01 using 0.001 sized bins and $a_c/a_{bin} =$ 0.01 -- 1.0 using 0.01 sized bins.  Figure \ref{fig:SType_prob} shows the PDF (Figs. \ref{fig:SType_prob}a--\ref{fig:SType_prob}c) and cumulative distribution function (CDF; \ref{fig:SType_prob}d--\ref{fig:SType_prob}f) resulting from the Monte Carlo integration considering a planetary inclination of $0^\circ$, $45^\circ$, and $180^\circ$.  For planets orbiting either the primary (Star A; black) or the secondary (Star B; gray) star, there are many binaries with $a_c/a_{bin} > 0.05$ and hence both curves in Figs. \ref{fig:SType_prob}a--\ref{fig:SType_prob}c have the highest probability within this range.  Planets that orbit the secondary constitute a larger fraction of binaries with a small critical semimajor axis and thus, the probability is slightly larger.  This trend reverses for $a_c/a_{bin} \gtrsim 0.12$ as the stability limit for the primary extends to larger distances compared to the secondary.  Sufficiently inclined planets (e.g., $i_p = 45^\circ$) have their stability limits truncated due to the large growth of eccentricity from the Lidov-Kozai mechanism and stable orbits are possible for predominantly low critical semimajor axis ratios.  Retrograde orbits extend to larger critical semimajor axis ratios \citep{Henon1970,Quarles2016} and the PDF is flatter.


The uncertainty in the binary semimajor axis $a_{bin}$ from observations can be quite large, where discoveries from direct imaging may only have an estimate for the projected semimajor axis.  From such situations, a CDF is more useful where the probability can be calculated from a difference of values on the CDF and a single value describes the fraction of binaries with a stability limit less than a given $a_c/a_{bin}$.  For example, Figure \ref{fig:SType_prob}d shows that half of all binaries within our domain will have a critical semimajor axis for coplanar planets that is less than 0.06 (Star B) or 0.078 (Star A).  The slope in the CDF for the Lidov-Kozai regime (Fig. \ref{fig:SType_prob}e) is steeper, which shifts the median to smaller $a_c/a_{bin}$.  Moreover, the slope is nearly identical between which star the planet is orbiting (star A or B).  The shallower slope for retrograde planets (Fig. \ref{fig:SType_prob}f) has a median critical semimajor axis closer to 0.1 and extends to larger values of $a_c/a_{bin}$.

The black and gray curves in Figs. \ref{fig:SType_prob}d--\ref{fig:SType_prob}f are well-fit by a modified exponential distribution:

\begin{equation}
    F(\xi) = 1 - e^{-c_1 \xi^2 - c_2 \xi},
\end{equation}

\noindent where $\xi = a_c/a_{bin}$.  The red curves in Figs. \ref{fig:SType_prob}d--\ref{fig:SType_prob}f mark sample curves within the 1$\sigma$ of the best-fit values for $F(\xi)$, where the largest differences occur in the top quintile as marked by the respective inset panels.  \cite{David2003} used a similar function, except their formulation depends on the binary periastron distance.  Table \ref{tab:CDF_coeff} provides values for the coefficients ($c_1$ and $c_2$) along with their uncertainties using the Levenberg-Marquardt algorithm in the \texttt{curve fit} function within the \texttt{scipy} module.  In addition to the coefficients, we provide the the critical semimajor axis values at 50$\%$ and 99.7$\%$ quantiles.

\subsection{Circumstellar Planets Observed with TESS}

NASA's TESS will observe nearly $85\%$ of the sky over the course of its primary two-year mission. The spacecraft has four cameras that provide a ${\rm 24^\circ \times 96^\circ}$ field-of-view, and a photometric precision between ${\rm \sim 100\ ppm}$ for apparent magnitude ${I_C\sim4}$, and ${\rm \sim 10^5\ ppm}$ for ${I_C\sim 18}$ \citep{Ricker2015}. It is expected that TESS will discover thousands of transiting exoplanets around bright, nearby stars \citep{Sullivan2015,Bouma2017,Barclay2018}, including hundreds of planets smaller than ${\rm 2R_\oplus}$. A significant number of stars (some of which will have transiting planets) observed by TESS will be in multiple systems, i.e.~from ${\rm \sim20\%\ for\ M_{prim} < 0.1\ M_\odot}$ to ${\rm \sim75\%\ for\ M_{prim} > 1.4\ M_\odot}$, and spanning semimajor axes from a few to hundreds of AU \citep{Sullivan2015}. In combination with data from Gaia, by virtue of being bright targets (and thus favorable to follow-up observations) those of the multiple star systems hosting S-type planet candidates will be amenable to measurements of the mass ratio. The results we present here will be directly applicable to the stability of these systems. 

Recently, \cite{Winters2019} uncovered a transiting planet in a triple star system, LTT 1445ABC, where star A ($M_A = 0.257$ M$_\odot$) orbits a nearly equal mass stellar binary ($M_B = 0.215$ M$_\odot$ and $M_C = 0.161$ M$_\odot$) and the separation from star A to the BC barycenter is $\sim$34 AU.  The nearly Earth-sized exoplanet LTT 1445Ab ($R_p \approx 1.38 R_\oplus$) orbits very quickly (5.358 days, or $a_p = 0.038$ AU) around LTT 1445A, which is well within the stability limit assuming that the orbit of LTT 1445A is nearly circular around the barycenter of LTT 1445BC.  However, the A-BC orbit is likely eccentric considering observations over the last few decades that indicate significant astrometric variation \citep[][and references therein]{Mason2009,Dieterich2012,Winters2019} and a non-circular orbit may affect the stability of the observed exoplanet.  Using the combined mass for LTT 1445BC ($M_{BC} = 0.376$ M$_\odot$), the dynamical mass ratio of the A-BC pair is $\mu = 0.594$ given that the planet is orbiting the less massive component of the pair.  The semimajor axis  $a_p$ is quite small (0.038 AU), where the planetary orbital stability may be compromised for a large eccentricity $e_{bin}$ of the A-BC orbit.

If we consider planets within the A-BC plane, then the stability limit varies between 1.38 AU -- 8.15 AU depending on the value of $e_{bin}$, the eccentricity of star A around the BC barycenter.  Even for a high value for $e_{bin}$, the stability limit is still $\sim36$ times farther from the host star than the observed planetary semimajor axis.  Therefore, we are confident that LTT 1445Ab is on a stable orbit as the planet is very far from the stability limit under reasonably extreme assumptions on the shape of the LTT 1445 A-BC stellar orbit.  If TESS observes the system during the extended mission and detects potential transits of additional planets on longer orbits, reducing the BC binary to its barycenter may need to be revised and the eccentricity of star A around the BC barycenter properly evaluated. We do not investigate non-planar conditions ($i_p =45^\circ$) where the stability limit can be further reduced because the astrometric measurements show that the LTT 1445ABC system is nearly planar and the planet is transiting, which implies that the planetary orbit is likely nearly aligned with the LTT 1445ABC orbital plane. 

\begin{deluxetable}{lcccc|cc|cc}
\tablecaption{Mass ratio, Semimajor Axis, and Stability Limits for Known Planet-hosting Binaries with $5\leq a_{bin} < 500$ au  \label{tab:binaries}}
\tablehead{ \colhead{System} & \colhead{$M_A$} & \colhead{$M_B$} & \colhead{$\mu$} & \colhead{$a_{bin}$} & \colhead{$a^{A,\dagger}_c$} & \colhead{$a^{A,\star}_c$} & \colhead{$a^{B,\dagger}_c$} & \colhead{$a^{B,\star}_c$} \\ 
& \colhead{($M_\odot$)} & \colhead{($M_\odot$)} && \colhead{(au)} & \colhead{($a_{bin}$)} & \colhead{($a_{bin}$)} & \colhead{($a_{bin}$)} & \colhead{($a_{bin}$)}}
\startdata
HD 109749 & 1.100 & 0.780 & 0.415 & 490 & 0.046 & 0.299 & 0.040 & 0.235 \\ 
HD 133131 & 0.950 & 0.930 & 0.495 & 360 & 0.043 & 0.267 & 0.042 & 0.263 \\ 
HD 106515 & 0.910 & 0.880 & 0.492 & 345 & 0.043 & 0.267 & 0.042 & 0.261 \\ 
Kepler-108 & 1.300 & 0.960 & 0.425 & 327 & 0.045 & 0.295 & 0.040 & 0.238 \\ 
WASP-77 & 1.002 & 0.710 & 0.415 & 306 & 0.046 & 0.299 & 0.040 & 0.235 \\ 
KELT-2 & 1.317 & 0.780 & 0.372 & 295 & 0.047 & 0.321 & 0.037 & 0.222 \\ 
HD 114729 & 0.930 & 0.253 & 0.214 & 282 & 0.053 & 0.379 & 0.031 & 0.169 \\ 
Kepler-14 & 1.510 & 1.390 & 0.479 & 280 & 0.043 & 0.272 & 0.042 & 0.257 \\ 
HD 27442 & 1.200 & 0.750 & 0.385 & 240 & 0.047 & 0.314 & 0.038 & 0.227 \\ 
TrES-2 & 0.980 & 0.509 & 0.342 & 232 & 0.048 & 0.338 & 0.037 & 0.213 \\ 
HD 212301 & 1.270 & 0.350 & 0.216 & 230 & 0.053 & 0.380 & 0.032 & 0.170 \\ 
HD 16141 & 1.010 & 0.286 & 0.221 & 220 & 0.052 & 0.380 & 0.032 & 0.171 \\ 
HD 189733 & 0.800 & 0.200 & 0.200 & 216 & 0.054 & 0.377 & 0.031 & 0.164 \\ 
HD 217786 & 1.020 & 0.160 & 0.136 & 155 & 0.057 & 0.435 & 0.027 & 0.141 \\ 
HD 142 & 1.200 & 0.590 & 0.330 & 138 & 0.048 & 0.346 & 0.036 & 0.209 \\ 
HD 114762 & 0.840 & 0.138 & 0.141 & 132 & 0.057 & 0.432 & 0.028 & 0.143 \\ 
HD 195019 & 1.060 & 0.700 & 0.398 & 131 & 0.046 & 0.307 & 0.040 & 0.230 \\ 
WASP-2 & 0.890 & 0.480 & 0.350 & 106 & 0.047 & 0.334 & 0.037 & 0.215 \\ 
HD 19994 & 1.340 & 0.350 & 0.207 & 100 & 0.054 & 0.378 & 0.031 & 0.167 \\ 
HD 177830 & 1.450 & 0.230 & 0.137 & 97 & 0.057 & 0.434 & 0.027 & 0.142 \\ 
Gliese 15 & 0.380 & 0.150 & 0.283 & 93 & 0.051 & 0.378 & 0.035 & 0.194 \\ 
Kepler-296 & 0.500 & 0.330 & 0.398 & 80 & 0.046 & 0.307 & 0.040 & 0.230 \\ 
GJ 3021 & 0.900 & 0.130 & 0.126 & 68 & 0.057 & 0.440 & 0.026 & 0.137 \\ 
K2-288 & 0.520 & 0.330 & 0.388 & 54.8 & 0.047 & 0.312 & 0.038 & 0.227 \\ 
HD 120136 & 1.400 & 0.400 & 0.222 & 45 & 0.052 & 0.380 & 0.032 & 0.172 \\ 
WASP-11 & 0.820 & 0.340 & 0.293 & 42 & 0.050 & 0.369 & 0.035 & 0.197 \\ 
K2-136 & 0.740 & 0.100 & 0.119 & 40 & 0.057 & 0.442 & 0.026 & 0.135 \\ 
HD 164509 & 1.130 & 0.420 & 0.271 & 37 & 0.051 & 0.383 & 0.034 & 0.190 \\ 
HD 41004 & 0.700 & 0.400 & 0.363 & 23 & 0.048 & 0.326 & 0.037 & 0.219 \\ 
HD 196885 & 1.330 & 0.550 & 0.293 & 23 & 0.050 & 0.369 & 0.035 & 0.197 \\ 
HD 4113 & 0.990 & 0.550 & 0.357 & 23 & 0.048 & 0.330 & 0.037 & 0.217 \\ 
GJ 86 & 0.800 & 0.490 & 0.380 & 21 & 0.047 & 0.317 & 0.038 & 0.225 \\ 
$\gamma$ Cep & 1.180 & 0.320 & 0.213 & 20 & 0.053 & 0.379 & 0.031 & 0.169 \\ 
HD 8673 & 1.400 & 0.400 & 0.222 & 10 & 0.052 & 0.380 & 0.032 & 0.172 \\ 
Kepler-420 & 0.990 & 0.700 & 0.414 & 5.3 & 0.046 & 0.300 & 0.039 & 0.235 \\ 
\enddata
\tablecomments{Critical semimajor axis $a_c$ calculated using interpolation of Fig. \ref{fig:interp_maps}a, where planets orbit star A for $\mu$ and star B for $1-\mu$.  The stellar parameters are from an online catalog of planets in binaries maintained by Richard Schwarz (\url{https://www.univie.ac.at/adg/schwarz/multiple.html}).}
\tablenotetext{\dagger}{$e_{bin}=0.8$}
\tablenotetext{\star}{$e_{bin}=0.0$}
\end{deluxetable}

\begin{deluxetable}{lccccc}
\tablecaption{Cumulative Distribution Function Coefficients for the Critical Semimajor Axis  \label{tab:CDF_coeff}}
\tablehead{\colhead{} & \colhead{$i_p$} & \colhead{$c_1 \pm \sigma_1$} & \colhead{$c_2 \pm \sigma_2$} & \colhead{$50\%$} & \colhead{$99.7\%$}}
\startdata
Star A & $0^\circ$ & $28.96 \pm 0.75$ & $6.65 \pm 0.09$ & 0.078 & 0.348\\
Star B & $0^\circ$ & $54.50 \pm 1.62$ & $8.31 \pm 0.14$ & 0.060 & 0.259\\
\hline
Star A & $45^\circ$ & $15.11 \pm 2.30$ & $13.28  \pm 0.23$ & 0.049 & 0.321\\
Star B & $45^\circ$ & $20.69 \pm 3.28$ & $13.82  \pm 0.30$ & 0.047 & 0.292\\
\hline
Star A & $180^\circ$ & $15.41 \pm 0.37$ & $4.82 \pm 0.06$ & 0.107 & 0.477\\
Star B & $180^\circ$ & $22.97 \pm 0.62$ & $5.88 \pm 0.08$ & 0.088 & 0.391\\
\enddata
\tablecomments{{The coefficients ($c_1$ and $c_2$) and uncertainties ($\sigma_1$ and $\sigma_2$)} using a fitting formula for the CDF, $F(\xi) = 1-e^{-c_1 \xi^2 - c_2 \xi}$ where $\xi = a_c/a_{bin}$. }
\end{deluxetable}

\section{Summary \& Discussion} \label{sec:conc}

The orbital stability of circumstellar planets within stellar binaries depends on many factors, where we focus on the binary mass ratio $\mu$, eccentricity $e_{bin}$, semimajor axis ratio $a_p/a_{bin}$, and the mutual inclination of the planet $i_p$.  \cite{Holman1999} developed a fitting formula that incorporates the first three of these parameters and has a limited accuracy (up to 11\%).  We update the coefficients of the fitting formula using a larger array of N-body integrations ($\sim$700 million simulations), where the larger dataset substantially narrows the uncertainties and calculates fitting coefficients that account for the planetary inclination $i_p$.  We provide additional formulations for stability using the Jacobi constant when $e_{bin}$ is small or a constraint on the maximum planetary eccentricity.  Using the known population of Sunlike stellar binaries, we find that the stability limit is typically less than $\lesssim$8\% of the binary semimajor axis $a_{bin}$, where this fraction can be substantially reduced for an inclined planet that is undergoing Lidov-Kozai cycles or expanded if the planet orbits in retrograde.

\cite{Eberle2008} developed a stability criterion based upon the value of the Jacobi constant at the collinear Lagrange points within the circular restricted three body problem (CRTBP).  We extend this criterion to include orbits around the less massive secondary star and with a significantly inclined planetary orbit.  However, this stability criterion has some limitations, where an asymmetry in the Coriolis acceleration \citep{Innanen1980,Hamilton1991,Grishin2017} and a reduction in the duration of action \citep{Henon1970} can promote greater stability.  The integration of the Jacobi integral (typically by quadrature) introduces a squared velocity term that is invariant for the direction of the smaller body's motion (i.e., planet) and forms a symmetry with respect to the planetary inclination.  We provide in Table \ref{tab:circ_stab} the binary mass ratios with the critical semimajor axis ratios considering the Jacobi constants ($C_1$ \& $C_3$) for prograde orbits and a fit to the stability boundary from our numerical simulations (Fig. \ref{fig:CRTBP}) for coplanar retrograde orbits ($C^{retro}_1$).  We note that for a special case in our results ($\mu=0.999$), the ratio of the retrograde critical semimajor axis ratio to the respective prograde ratio agrees well with the analytical expectation (i.e., $C^{retro}_1/C_1 \approx 3^{1/3}$) derived for hierarchical three-body systems \citep{Hamilton1991,Grishin2017}.  Although highly inclined orbits experience a greater eccentricity excitation, they also feel a smaller contribution from the Coriolis force and can thereby achieve capture into mean motion resonances with the binary \citep{Morais2012,Morais2013}.

We provide updated coefficients ($c_1 - c_6$) to the stability formulas by \cite{Holman1999}, where the uncertainties of these coefficients are typically reduced by a factor of 3.  Moreover, we calculate coefficients using numerical simulations for a range of planetary inclination ($i_p = 0^\circ$, $30^\circ$, $45^\circ$, and $180^\circ$).  For nearly coplanar planets in prograde, the coefficients have similar mean values.  The coefficients for prograde planets undergoing Lidov-Kozai oscillations ($i_p = 45^\circ$) are more strongly dependent on the binary eccentricity.  The coefficients fitting our numerical results for coplanar, retrograde planets depend linearly on the binary eccentricity, in contrast to the prograde results.  Our coefficients are more accurate, but suffer from errors due to averaging similar to \cite{Holman1999}.  Therefore, we provide interpolation maps, where the data behind these maps are available to the community through \dataset[GitHub]{https://github.com/saturnaxis/ThreeBody_Stability} and \dataset[Zenodo]{http://doi.org/10.5281/zenodo.3579202} that is similar to our previous repositories \citep{Quarles2018b}.  In the \texttt{GitHub} repository, there is an example script in Python that demonstrates how to use the interpolation schemes in \texttt{scipy} to estimate the stability limit between grid values.  Using our interpolation scheme, the stability limit for a prograde, coplanar planet orbiting $\alpha$ Cen A or $\alpha$ Cen B is 2.78 AU or 2.60 AU, respectively, and is consistent with a previous detailed billion year study of the system \citep{Quarles2018a}.  However a retrograde, coplanar planet orbiting $\alpha$ Cen A or $\alpha$ Cen B has a larger stability limit of 3.84 AU or 3.61 AU, respectively, and can be extended farther when the eccentricity vectors of the planetary and binary orbits begin in a relative alignment that minimizes the free eccentricity \citep{Quarles2018a}.

Population studies of Sunlike binary stars show that the mass quotient $q$, binary eccentricity $e_{bin}$, and binary period $P_{bin}$ can be statistically fit using either power laws or a log normal distribution \citep{Raghavan2010,Moe2017}.  We use these statistical fits to estimate the probability of a planet (via a PDF or CDF) orbiting star A rather than star B given the critical semimajor axis ratio $a_c/a_{bin}$.  The PDF (or CDF) is modified depending on the assumed mutual inclination of the planet where: a coplanar prograde planet has the most probability when $a_c/a_{bin} \lesssim 0.08$, an inclined ($i_p = 45^\circ$) prograde planet's high probability regime decreases by half ($a_c/a_{bin} \lesssim 0.04$), and a coplanar retrograde planet has a broader high probability region extending to $a_c/a_{bin} \lesssim 0.10$.  Table \ref{tab:CDF_coeff} provides coefficients fitted to each CDF using a modified exponential distribution.  Using $a_c/a_{bin}$ is important for future studies because direct imaging observations often rely on projected distances and new disk observations of young binary systems can match this ratio to the disk truncation distance \citep[e.g.,][]{Alves2019}. 

Observations are ongoing to find and characterize circumstellar planets in binary star systems, where the TESS mission as well as Gaia \citep{Gaia2018} will be instrumental for future discoveries.  In particular, \cite{Winters2019} identified a planet within a hierarchical triple star system, LTT 1445ABC, using data from TESS.  Due to the long orbital binary orbital periods necessary for stability, Gaia will uncover the multiplicity of stellar systems \citep[e.g.,][]{Evans2018}, while TESS may identify whether any planets are transiting at the present epoch.  Additional efforts will soon be underway using the James Webb Space Telescope (JWST), where direct imaging help identify the extent of the protoplanetary disks within young stellar binaries or whole systems of planets in older systems.  \cite{Beichman2019} outlined particular considerations needed to use JWST for this purpose in observing $\alpha$ Cen A and some \textit{a priori} knowledge for the stability of planets is beneficial for such observations.

\acknowledgments
Some of the computing for this project was performed at the OU Supercomputing Center for Education \& Research (OSCER) at the University of Oklahoma (OU).  This research was supported in part through research cyberinfrastructure resources and services provided by the Partnership for an Advanced Computing Environment (PACE) at the Georgia Institute of Technology.  NH acknowledges support from NASA XRP program through grant \#80NSSC18K0519.

\software{scipy \citep{Virtanen2019}; mercury6 \citep{Chambers2002}; REBOUND \citep{Rein2012,Rein2015}}
\newpage
\bibliography{bibliography}
\bibliographystyle{aasjournal}


\end{document}